\documentclass[12pt]{article}

\RequirePackage{graphicx}
\usepackage{xcolor}
\usepackage{scicite}
\usepackage{times}

\usepackage{amsmath,amstext,amssymb}
\usepackage{wasysym} 						
\usepackage{verbatim}
\usepackage{soul}
\usepackage[T1]{fontenc}
\newcommand{\approptoinn}[2]{\mathrel{\vcenter{
 \offinterlineskip\halign{\hfil$##$\cr
  #1\propto\cr\noalign{\kern2pt}#1\sim\cr\noalign{\kern-2pt}}}}}
\newcommand{\appropto}{\mathpalette\approptoinn\relax}
\newcommand{\be}{\begin{equation}}
\newcommand{\ee}{\end{equation}}
\def\bs{\boldsymbol}

\newenvironment{sciabstract}{%
\begin{quote} \bf}
{\end{quote}}

\title{Understanding Density  Fluctuations in Supersonic, Isothermal Turbulence}

\author{
Evan Scannapieco,$^{1\ast}$ Liubin Pan,$^{2 \ast}$ Edward Buie II$^{3}$, and Marcus Br\" uggen$^{4}$\\
\\
\normalsize{$^{1}$School of Earth \& Space Exploration, Arizona State University,}\\
\normalsize{781 Terrace Mall, Tempe, AZ 85287, USA}\\
\normalsize{$^{2}$School of Physics and Astronomy, Sun Yat-sen University,}\\
\normalsize{2 Daxue Road, Zhuhai, Guangdong, 519082, People's Republic of China}\\
\normalsize{$^{3}$Physics \& Astronomy Department, Vassar College,}\\
\normalsize{124 Raymond Avenue, Poughkeepsie, NY 12604, USA}\\
\normalsize{$^{4}$Universit\"at Hamburg,}\\
\normalsize{Hamburger Sternwarte, Gojenbergsweg 112, 21029, Hamburg, Germany}\\
\normalsize{$^\ast$ Corresponding authors: evan.scannapieco@asu.edu, panlb5@mail.sysu.edu.cn}
}

\date{}

\begin{document}

\baselineskip15pt


\maketitle 

\begin{sciabstract}

Abstract: {\normalfont  Supersonic turbulence occurs in many environments, particularly in astrophysics. In the crucial case of isothermal turbulence, the probability density function (PDF) of the logarithmic density, $s$, is well measured, but  a theoretical understanding of the processes leading to this distribution remains elusive. We investigate these processes using Lagrangian tracer particles to track $s$ and $\frac{ds}{dt}$ in direct numerical simulations, and we show that their evolution can be modeled as a stochastic differential process with time-correlated noise. The temporal correlation functions of $s$ and $\frac{ds}{dt}$ decay exponentially, as predicted by the model, and the decay timescale is $\approx$ 1/6 the eddy turnover time. The behavior of the conditional averages of $\frac{ds}{dt}$ and $\frac{d^2s}{dt^2}$  is also well explained by the model, which shows that the density PDF arises from a balance between stochastic compressions/expansions, which tend to broaden the PDF, and the acceleration/deceleration of shocks by density gradients, which tends to narrow it.}\\

Teaser: {\normalfont Innovative numerical and analytic techniques yield deeper insights into the density statistics of supersonic turbulence.}

\end{sciabstract}


\section{Introduction}

Supersonic turbulence occurs in a wide variety of contexts including supersonic flight \cite{Chang18}, volcanic plumes \cite{Ogeden08}, laboratory experiments \cite{White19,Liao19}, and many astrophysical systems \cite{Brandenburg95,Schekochihin09,Moesta15,Walch15,Kim17,Buie20}. While fully understanding such systems often requires tracking magnetic fields \cite{Zweibel95, Li04, Li15, Xu19, Seifried20,Kim21,Pattle22}, chemical reactions \cite{vanDishoeck98, Glover12, Ramirez13}, and other processes  \cite{Wang10, Girichidis16, Kim18}, many of their overall features can be probed by considering the fundamental case of hydrodynamic turbulence with a simplified isothermal equation of state \cite{Larson81,Federrath10}.  Moreover, supersonic, isothermal turbulence is an essential starting point for understanding giant molecular clouds, where all stars are born, and whose properties set the stellar initial mass function and star formation rate  \cite{MacLow04,Federrath12,Padoan14}.

In such turbulence, energy injection by a large-scale stochastic acceleration field cascades toward small scales,and shocks and complex density structures form through highly nonlinear processes. In continually-forced turbulence, the probability density function (PDF) of density fluctuations  achieves a global steady state, even though individual parcels of gas are constantly changing. In this case, the mass-weighted PDF is well approximated by a Gaussian
\be
P_{\rm M}(s) \approx \frac{1}{\sqrt{2 \pi \sigma_s^2}} {\rm exp} \left[ - \frac{(s - \left<s\right>)^2} {2 \sigma_s^{2}} \right],
\label{eq:PDF}
\ee
where $s \equiv \ln(\rho/\rho_0),$ where $\rho_0$ is the mean density, and the mean value of $s$ is related to the variance, $\sigma_s^2$, by $\left<s\right> = \sigma_s^2/ 2$, as required by mass conservation \cite{VazquezSemadeni94,Padoan97,Federrath08,Federrath10,Padoan11}. Previous studies have established that $\sigma_s^2$ can be approximately fit by a simple function of the Mach number \cite{Padoan97, Ostriker2001, Price11, MacLow05, Kowal07, Glover07, Lemaster08}. These have also shown that at high Mach numbers and/or with compressive driving, $P_{\rm M}(s)$ becomes skewed towards the low-density tail \cite{Kritsuk07, Federrath08, Burkhart09, Schmidt09, Federrath10, Konstandin12, Hopkins13, Federrath13, Squire17, Pan19}.

While the structure and lifetime of dense regions have been the subject of previous investigations \cite{Falceta11, Robertson18}, the evolution of $s$ along Lagrangian trajectories has been poorly constrained. 
Nevertheless in astrophysical environments, it is this evolution that controls which parcels of gas will collapse due to gravity and which will be reshuffled to lower densities \cite{Krumholz05, Padoan11, Hennebelle11, Federrath12, Hopkins13}, see however \cite{Elmegreen00, Murray11}.  Recently, \cite{Scannapieco18} proposed that density fluctuations in the Lagrangian frame evolve according to a Langevin model, a differential equation that defines how $s$ evolves under a combination of stochastic and deterministic forces. By using a passive scalar field they were able to test this model by measuring the change in $s$ over a limited number of time intervals. This approach was extended in \cite{Mocz19}, which focused on the skewness of the PDF.

Here we return to the question of the evolution of density fluctuations in direct numerical turbulence simulations and constructing an analytic model that captures this evolution. 

\begin{table*}[t]
 \centering
 \resizebox{1.0\textwidth}{!}{
\begin{tabular}{|l|c|c|c|c|c|c|c|c|c|}
\hline
\, \, Name & $M_{\rm rms}$ & $M_{\rm ave}$ & $\tau_{\rm e} $ &  $\overline{\Delta t} $ & $\nu_0 $ & $\nu$  & \, Re \, \\
\, \,  &  & & $(\tau_{\rm sc})$ & $(\tau_{\rm sc})$  & $(L_{\rm box} c_s)$ & $(L_{\rm box} c_s)$ &  \\
\hline
M3$\nu_0$0   & 3.2 & 2.9 & 0.16   & $8.9 \times 10^{-5}$ & 0   & $3.1 \times 10^{-4}$   & 5,100    \\
M3$\nu_0$1   & 3.2 & 2.9 & 0.16   & $9.1 \times 10^{-5}$ & $1.0 \times 10^{-4}$   & $4.3 \times 10^{-4}$   & 3,700     \\
M3$\nu_0$3   & 3.2 & 2.9 & 0.16   & $9.3 \times 10^{-5}$ & $3.0 \times 10^{-4}$   & $6.1 \times 10^{-4}$   & 2,600  \\
M3$\nu_0$6   & 3.2 & 2.9 & 0.16   & $9.5 \times 10^{-5}$ & $6.0 \times 10^{-4}$   & $8.5 \times 10^{-4}$   & 1,900   \\
M1.3$\nu_0$3  & 1.3 & 1.2 & 0.38   & $20 \times 10^{-5}$ & $3.0 \times 10^{-4}$   & $3.6 \times 10^{-4}$   & 1,800    \\
M2$\nu_0$3   & 1.8 & 1.7 & 0.27   & $15 \times 10^{-5}$ & $3.0 \times 10^{-4}$   & $5.1 \times 10^{-4}$   & 1,800   \\
M4$\nu_0$3   & 4.3 & 3.9 & 0.11   & $5.5 \times 10^{-5}$ & $3.0 \times 10^{-4}$   & $6.2 \times 10^{-4}$   & 3,500   \\
M6$\nu_0$3   & 5.9 & 5.4 & 0.085  & $3.4 \times 10^{-5}$ & $3.0 \times 10^{-4}$   & $7.0 \times 10^{-4}$   & 4,200   \\
\hline 
\end{tabular}}
\vspace{0.1in}
\caption{Parameters of our simulations. Columns show the run name, mass-weighted rms and average Mach numbers, eddy turnover time and average timestep in units of the box sound crossing time, explicit and effective viscosity in units of the box size and sound speed, and the effective Reynolds number.}
\label{tab:runs}
\end{table*}

\section{Results}

\subsection{Simulation Suite}

Our study builds on what has become a standard approach to produce a suite of simulations of supersonic, isothermal turbulence. As described in more detail in the Materials and Methods section,  we carried out each simulation in a $512^3$ periodic box of size $L_{\rm box},$ over which we solved the hydrodynamic equations in the presence of a stochastic driving acceleration, $\bs a,$ that is  solenoidal  ( $\nabla \cdot \bs a = 0$).
We drove the flows with the same power at all Fourier modes with $1/L_{\rm box} \le|\bs k|/2 \pi \le 3/ L_{\rm box} $. Several authors have used Lagrangian tracer particles to model shock dynamics \cite{Biferale05, Yang13, Toschi09}, and we adopted a similar approach to track the Lagrangian evolution of density, with each simulation including $128^3$ tracer particles that evolved passively and retained the values of $s$ and $\frac{ds}{dt}$ at a range of time intervals.  In this initial study, we only consider turbulence with solenoidal driving, which is of fundamental theoretical interest. Although a mixture of compressive ($ \nabla \times \bs{a } =  0$) and solenoidal driving may provide a better description of some, but not all, astrophysical systems, \cite[e.g.]{Federrath16}, but it is more challenging to interpret as the driving force may act directly on $P_M(s)$ \cite{Pan19}.

By varying the strength of the driving and the explicit viscosity, we carried out eight simulations, spanning a wide parameter range.  Here we focus purely on shear viscosity, although bulk viscosity may also be  a consideration in compressible flow \cite{Beattie23}. The runs can be divided into two sets. In the first set, we fixed the amplitude of the driving term, such that the flow reached a steady state with a mass-weighted rms Mach number of $M_{\rm rms} = 3.2,$ and we varied the value of the explicit viscosity, $\nu_0,$ between 0 and $6\times10^{-4} L_{\rm box} c_s$ (or 0.31 $\Delta x c_s$ where $\Delta x$ is the cell size and $c_s$ the sound speed). In the second case, we held the explicit viscosity constant at $\nu_0 = 3\times10^{-4} \, L_{\rm box} c_s$ and varied the forcing such that the Mach number ranged between $M_{\rm rms} =$ 1.3 and 5.9.  Each run reached a steady state after two to three eddy turnover times
\cite{Kritsuk07,Federrath10}, defined as 
\be
\tau_{e} \equiv \frac{\ell}{M_{\rm rms} c_{s}},
\ee
where $\ell$ is the driving scale of the turbulence.  After this period, the Mach number fluctuated within $\lesssim10\%$, and we let each simulation run for $\approx 15 \, \tau_e$ to compile accurate statistics.
 
Table \ref{tab:runs} summarizes the parameter space spanned by our runs. Each run is named according to its Mach number and explicit viscosity value. Here the mass-weighted root-mean-square (rms) Mach number and average Mach numbers $M_{\rm rms}$ and $M_{\rm ave}$ are given in columns 2 and 3. Also given in this table are the large eddy turnover time (in column 4) and the average time step in the steady-state portion of each run (in column 5).
 
 \begin{figure}[t]
\hspace{0.15in} \includegraphics[width=0.96\textwidth]{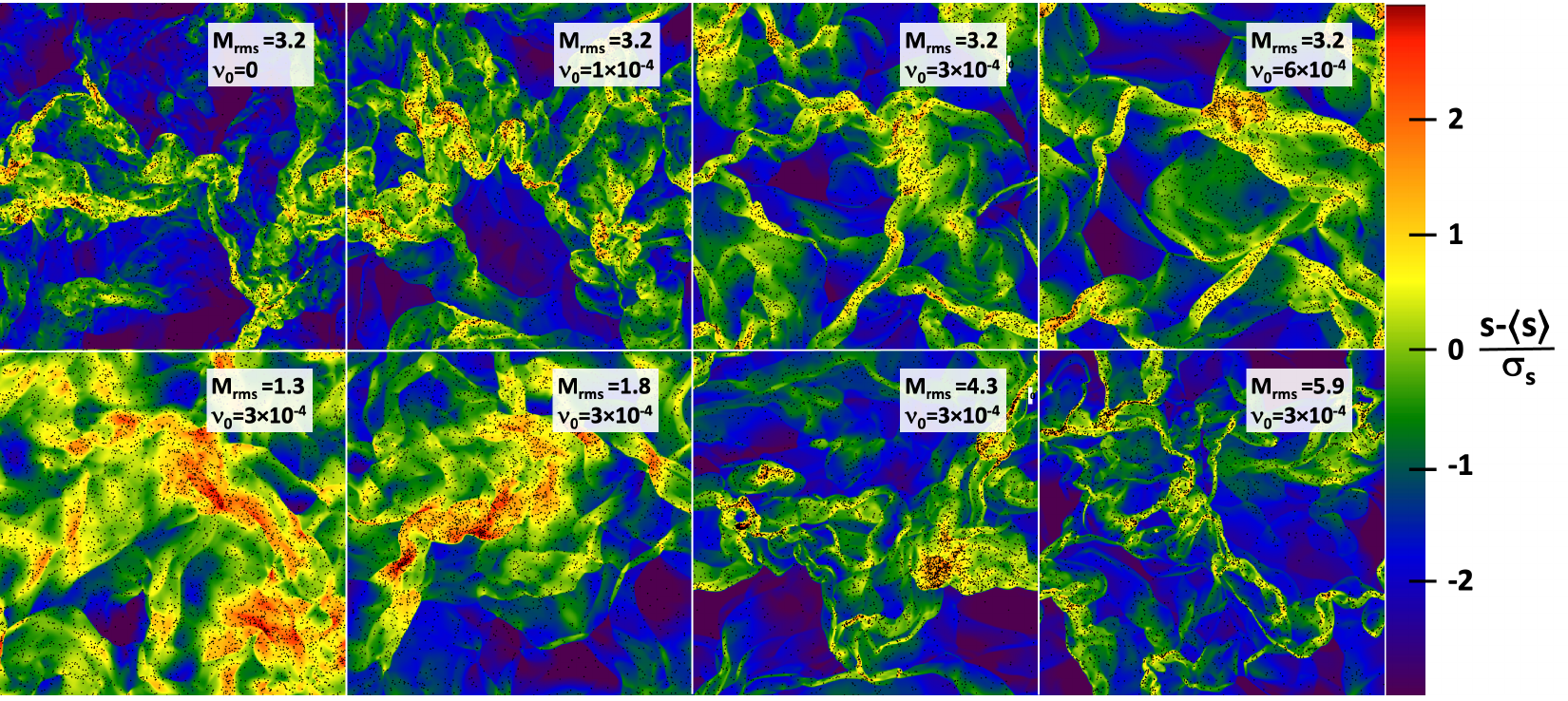}
\vspace{-0.1in}
\caption{{\bf Slices of $s-\left<s\right>$ from our turbulence simulations.} Each panel is  normalized by $\sigma_s$ and labeled by the mass-weighted Mach number and explicit viscosity. The top row shows the impact of changing $\nu_0$ at a fixed $M_{\rm rms}$, while the lower row shows the impact of changing $M$ at a fixed $\nu_0.$ The color bar is chosen to go from $-3 \sigma_s$ to $3 \sigma_s$ in all panels, with $\sigma_s$ = 0.98, 1.00, 1.01, and 1.03, across the top row, and $\sigma_s$ = 0.45, 0.68, 1.16, and 1.30 across the bottom row. Tracer particles with positions within $1\% L_{\rm box}$ above or below the slice are shown as the superimposed black points.}
\label{fig:slices}
 \end{figure}
 
In column 6, we give the value of the explicit viscosity in each of our runs. Note that even in the absence of an explicit viscosity, numerical diffusion leads to a significant effective viscosity \textbf{\cite{Malvadi23}}.  As detailed in the Materials and Methods section, we evaluate the effective viscosity according to the approach described in \cite{Pan10} and report its value as $\nu$ in column 6. Finally, the effective Reynolds number, $M_{\rm rms} c_s \ell / \nu,$ is given in column 7, {\rm yielding values consistent with previous studies} \cite{Malvadi23}.
 
Fig.\ \ref{fig:slices} shows slices of $s$ taken from each of our simulations at a time at which the turbulence is fully developed. To emphasize the features in each panel, we vary the color scale to span a range from $\left< s \right> \pm 3 \sigma_s$ from each simulation, using the values listed in Table \ref{tab:sigmas} below. The upper panels of this figure show the impact of changing $\nu_0.$ As viscosity increases, fewer small-scale structures are seen, because the velocity cascade is truncated at larger scales. Another more subtle effect is the smearing that occurs as shock fronts become more extended at higher $\nu_0$ values. 

The lower panels in this figure show the impact of the Mach number. As $M_{\rm rms}$ increases, $\sigma_s$ increases, and thus the material at high $s$ values is concentrated into a much smaller volume fraction. This is also true for the tracer particles, which can be seen spread throughout the volume when $M_{\rm rms}=1.3$, but strongly clustered when $M_{\rm rms}=5.9.$ Because these particles retain their histories, they allow us to track the Lagrangian statistics of the density fluctuations.

\subsection{Probability Distribution Functions}
\label{sec:pdf}
\begin{figure}[t]
\begin{center}
\includegraphics[width=1.0\textwidth]{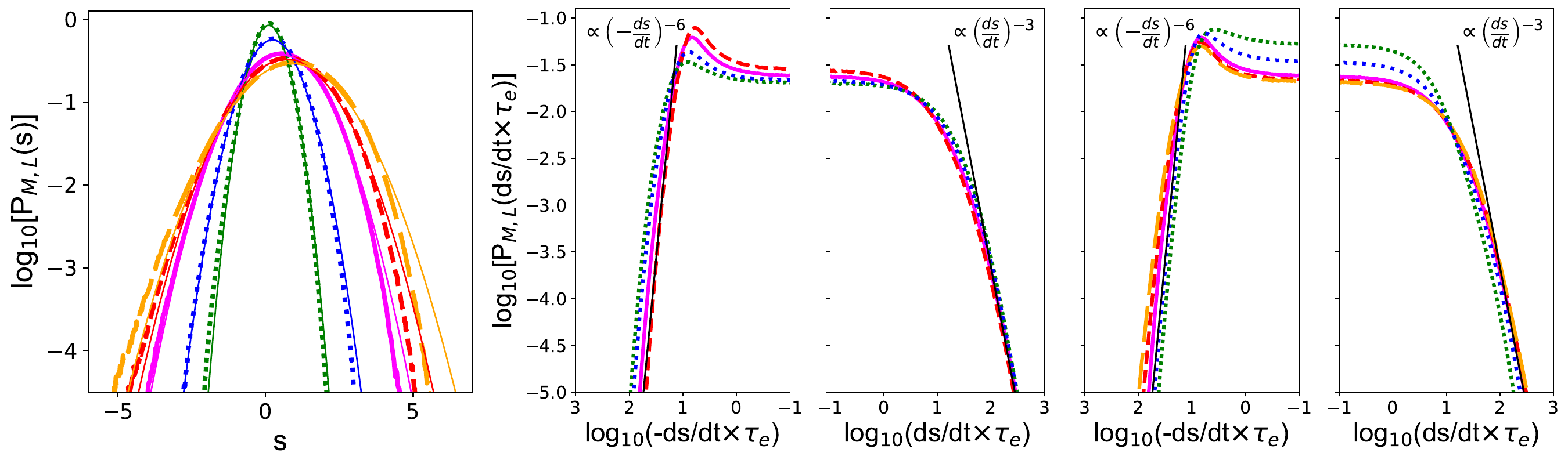}
\end{center}
\vspace{-0.2in}
\caption{{\bf  Mass-weighted  probability density function (PDF) as measured from the Lagrangian particles.} {\em Left:} Mass-weighted PDF of $s$ from $\nu_0= 3\times10^{-4} L_{\rm box} c_s$ simulations with varying Mach numbers. The colored lines show $P_{M,L}(s)$ for M1.3$\nu_0$3 (green dense-dotted), M2$\nu_0$3 (blue dotted), M3$\nu_0$3 (magenta solid), M4$\nu_0$3 (red short-dashed), and M6$\nu_0$3 (orange long-dashed). The thin solid lines show Gaussian PDFs with the same $\langle s \rangle$ and $\sigma_s^2$ as the measured $P_{M,L}(s)$, highlighting the negative skewness of the distributions. {\em Center:} The probability distribution, $P_{M,L}$ of $\frac{ds}{dt}$   measured from tracer particles in $M_{\rm rms}=3.2$ simulations as a function of $\nu_0.$ The colored lines show $P_{M,L}(\frac{ds}{dt} \times \tau_e)$ for M3$\nu_0$0 (green dense-dotted), M3$\nu_0$1 (blue dotted), M3$\nu_0$3 (magenta solid), and M3$\nu_0$6 (red short-dashed). To bring out the features of this PDF, the $x$-axis is divided into two log intervals. Note however that this is the PDF of $\frac{ds}{dt} \times \tau_e$ and not the PDF of $\log(\frac{ds}{dt} \times \tau_e).$  {\em Right:} $P_{M,L}(\frac{ds}{dt} \times \tau_e)$ for the $\nu_0= 3\times10^{-4} L_{\rm box} c_s$ simulations as function of Mach number. The colored lines show $P_{M,L}(\frac{ds}{dt} \times \tau_e)$  for M1.3$\nu_0$3 (green dense-dotted), M2$\nu_0$3 (blue dotted), M3$\nu_0$3 (magenta solid), M4$\nu_0$3 (red short-dashed), and M6$\nu_0$3 (orange long-dashed). The black lines are $\propto (-\frac{ds}{dt})^{-6}$ on the left and $\propto (-\frac{ds}{dt})^{-3}$, illustrating the overall asymmetry in the tails of $P_{M,L}(\frac{ds}{dt}\times \tau_e).$}
\label{fig:PsPdsdt}
\end{figure}

In the left panel of Fig.\ \ref{fig:PsPdsdt}, we show the probability distribution of $s$ for our $\nu_0= 3\times10^{-4} L_{\rm box} c_s$ simulations as a function of Mach number. 
Based on the ergodic  theorem for a statistically stationary system and the fact that the frequency at which tracer particles visit a particular region scales linearly with density, the PDF as measured by tracers is equivalent to the PDF with density weighting, as so we denote such measurements with a subscript {M,L} to indicate that are mass-weighted, as traced by Lagrangian particles.  $P_{M,L}(s)$ is always close to a Gaussian, with a variance that increases with Mach number \cite{Padoan97, Passot98, Klessen00, MacLow05, Kowal07, Kritsuk07, Glover07, Lemaster08, Federrath08, Federrath10,Pan10, Price11, Pan19}. We give the mean and variance of $s$ for each of these simulations in columns 2 and 3 of Table \ref{tab:sigmas}. In the Gaussian case, these quantities are related by mass conservation as $\langle s \rangle = \sigma_s^2/2$, and this relation is nearly (but not exactly) followed throughout our simulation suite. 

As in previous solenoidally-driven simulations, the variance of $P_{M,L}(s)$ is approximately fit by $\sigma_s^2 \approx \ln(1+b^2 M_{\rm rms}^2)$, and $b \approx 0.4$ \cite{Padoan97, Ostriker2001, MacLow05, Kowal07, Glover07, Lemaster08, Price11, Molina12}. Unlike the Mach number, the impact of viscosity on the $P_{M,L}(s)$ is extremely weak,  consistent with previous results that $P_{M}(s)$ already converges at relatively low resolution \cite{Kritsuk07, Federrath08, Burkhart09, Schmidt09, Federrath10, Konstandin12, Hopkins13, Federrath13, Squire17, Pan19}.  For this reason, we do not plot $P_{M,L}(s)$ from runs with varying $\nu$ values in Fig.\ \ref{fig:PsPdsdt}. 

\begin{table*}[t]
 \resizebox{1.0\textwidth}{!}{
\begin{tabular}{|l|c|c|c|c|c|c|c|c|c|c|c|}
\hline
\, \, Name & $\langle s \rangle$ & $\sigma^2_s$ & $\mu_s$ & $F_{{ds}/{dt}^+}$ & $\langle \frac{ds}{dt} \rangle \tau_e$ &  $\sigma_{{ds}/{dt}}^2 \tau_e^2$ & $\sigma_{{ds}/{dt}^+}^2/ \sigma_{{ds}/{dt}}^{2}$& 
 $\langle \frac{ds}{dt} \rangle \tau_e$ &  $\sigma_{{ds}/{dt}}^2 \tau_e^2$ \\
\, \, 	   &    		& \,   & & (Lag.)  & (Lag.) & (Lag.)  & (Lag.) & (Eul.)  & (Eul.) \\
\hline
M3$\nu_0$0 & 0.49 & 0.96 & -0.060 & 0.34 & 0.0078 & 1,050 & 0.82 & 2.2 & 800 \\
M3$\nu_0$1    & 0.50 & 1.00 & -0.077 & 0.31 & 0.0077 & 920  & 0.85 & 1.8 & 670 \\
M3$\nu_0$3  & 0.52  & 1.03 & -0.063 & 0.28 & -0.0017 & 720  & 0.88 & 1.4 & 540 \\
M3$\nu_0$6   & 0.54 & 1.06 & -0.074 & 0.27 & -0.00035 & 540  & 0.89 & 1.0 & 410 \\
M1.3$\nu_0$3   & 0.10 & 0.20 & -0.092  & 0.32 & -0.0016 & 190 & 0.84 & 0.094 & 170 \\
M2$\nu_0$3    & 0.23 & 0.46 & -0.052 & 0.30 & 0.0024 & 410  & 0.87 & 0.32 & 350 \\
M4$\nu_0$3    & 0.69 & 1.35 & -0.100 & 0.29 & 0.00039 & 920  & 0.94 & 2.4 & 660 \\
M6$\nu_0$3   &0.89  & 1.68 & -0.150 & 0.29 & -0.00033 & 1,120  & 0.97 & 3.9 & 700 \\
\hline 
\end{tabular}}
\vspace{0.1in}
\caption{Properties of our simulations. Columns show the run name, the mean, variance, and skewness of $P_{M,L}(s)$ as computed from the particles, the fraction of particles undergoing compressions, the mean, variance of $P_{M,L}(\frac{ds}{dt})$ and the positive contribution to the variance as computed from the particles, and the mean and variance of $P_{\rm M}(\frac{ds}{dt})$  as computed from the divergence of the velocity field.}
\label{tab:sigmas}
\end{table*}

Also shown in column 4 of Table \ref{tab:sigmas} is the skewness of $P_{M,L}(s),$ defined as $\mu \equiv \left< (s - \langle s \rangle)^3 \right>/\sigma_s^3.$ Negative (or positive) values of $\mu$ indicate that the tails of $P_{M,L}(s)$ are biased towards low (or high) $s$ values.
The skewness of our full simulation set is small and negative, as seen in previous solenoidally-driven simulations spanning this range of Mach numbers \cite{Federrath13, Pan19}. To illustrate this skewness in the left panel of Fig.\ \ref{fig:PsPdsdt} we also plot the Gaussian PDFs with the $\langle s \rangle$ and $\sigma_s^2$ as the measured.

The central and right panels of Fig.\ \ref{fig:PsPdsdt} show the PDF of $\frac{ds}{dt}$, i.e., the rate of change in $s$, 
which we denote as $P_{M,L}(\frac{ds}{dt})$, as it is computed by averaging over Lagrangian particles.{\ 
Here and below we express time in units of the eddy-turnover time, working with the dimensionless quantity 
$\frac{ds}{dt} \times \tau_e$. Unlike $P_{M,L}(s),$ $P_{M,L}(\frac{ds}{dt})$ is highly nongaussian, and strongly peaked near small values. 
To emphasize the features in this case, we adopt a log $x$-axis, dividing our plot into negative and positive ranges.

The resulting plot shows that $P_{M,L}(\frac{ds}{dt})$ is asymmetric, with a significant peak to the left of $\frac{ds}{dt}=0$, corresponding to the fact 
that the majority of the mass in the turbulent medium is slowly expanding. 
We integrate the right half of the histogram to compute the compressing mass fraction as $F_{{ds}/{dt}^+} \equiv \int_0^{\infty} d (\frac{ds}{dt}) P_{M,L}(\frac{ds}{dt}), $ which is given in the 5th column of Table \ref{tab:sigmas}. In all cases, one-third or less of the mass is being compressed.

To maintain a steady state, the average value of $\frac{ds}{dt}$ must be zero. This value is given in column 6 of Table \ref{tab:sigmas}, which shows that  
 $\left< \frac{ds}{dt} \right>$ is essentially zero in comparison with the rms of $\frac{ds}{dt}$ as given in column 7. A direct consequence of this is that the average value of $\frac{ds}{dt}$ on the right side of the histogram must be greater than on the left, meaning that while rarer than expanding zones, compressing zones undergo more rapid changes. This is also evident from the shape of the PDF, which drops more sharply on the negative side than on the positive side. This can be quantified by power-law fit to the tails of $P_{M,L}(\frac{ds}{dt})$, which shows that on the negative side $P_{M,L} \appropto(-\frac{ds}{dt})^{-6},$ while on the positive side $P_{M,L} \appropto(\frac{ds}{dt})^{-3}.$ 

This asymmetry means that the variance $\sigma_{{ds}/{dt}}^2$ is mainly contributed by particles with positive  $\frac{ds}{dt}$, which is indeed observed in column 8, which shows the fraction of $\sigma_{{ds}/{dt}}^2$ from the positive side of the PDF.  We find that, especially at high Mach numbers, almost all of the variance of $\frac{ds}{dt}$ arises from compressions,   in particular from shocks.

Unlike $\sigma_s^2$, the total variance of $P_{M,L}(\frac{ds}{dt}),$ is a strong function of both viscosity and Mach number. As $\nu$ increases, 
$\sigma_{{ds}/{dt}}^2$ drops monotonically, suggesting that compressions are slower at larger viscosities. On the other hand, 
$\sigma_{{ds}/{dt}}^2$ and $\sigma_s^2$ both increase with increasing $M_{\rm rms}$, but the dependence of $\sigma_{{ds}/{dt}}^2$ with Mach number is slightly weaker 
than that of $\sigma_s^2.$ Note however that since the {\em effective} viscosities of these runs slowly increase with Mach number, the difference in scaling could be due to the 
slowing of compressions with viscosity.

From the continuity equation (eq.\ \ref{eq:continuity}), we have through a change of variable
\be
\frac{d s}{d t} = \frac{\partial s}{\partial t} +  v_i \frac{\partial s}{\partial x_i} = - \frac{\partial v_i}{\partial x_i}.
\label{eq:dstoV}
\ee 
Therefore, as mentioned earlier, the Lagrangian PDF can be expressed as the density weighted PDF of $-\frac{\partial v_i}{\partial x_i}$ in the Eulerian frame, i.e., $P_{M,L}(\frac{ds}{dt}) = \langle \rho/\rho_0 \delta(\frac{ds}{dt} +\frac{\partial v_i}{\partial x_i})\rangle_E$, where the weighting factor $\rho/\rho_0$ is the ratio of the local density to the mean density.  
Here and below, for clarity of notation, we use the ensemble average $\langle Q \rangle$ to denote the average of any quantity $Q$ in the Lagrangian frame, whereas we denote the average in the Eulerian frame as $\langle Q \rangle_E$ with a subscript ``E.'' The two averages are related by $\langle Q \rangle = \langle \rho/\rho_0 Q \rangle_E$.

With this notation, the average of $\frac{ds}{dt}$ in the Lagrangian frame is equal to $-\langle  \rho/\rho_0 \nabla \cdot \bs v \rangle_E$, and is thus related to the $pdV$ work as 
$ \langle \frac{ds}{dt} \rangle = -\langle p \nabla \cdot \bs v \rangle_E/( \rho_0 c_s^2)$ in an isothermal flow. Therefore, like $\langle \frac{ds}{dt} \rangle,$ the total $pdV$ work should be 
equal to zero, as proven by \cite{Pan19}. However, column 9 shows that the Eulerian average $-\langle   \rho/\rho_0 \nabla \cdot \bs v \rangle_E$ is significantly nonzero. This is because,  
to enforce numerical stability, the Riemann solver needs to artificially modify the flow profiles around shocks. 
Therefore, as pointed out in \cite{Pan18, Pan19}, computing the statistics of the velocity divergence from the Eulerian grid is not sufficiently accurate to study $\frac{ds}{dt}$ in supersonic turbulence. 
Spatial gradients can be reliably evaluated from the grid only if the explicit viscosity is high enough for the shocks to be fully resolved. 
Note, however, that the overall $\sigma^2_{{ds}/{dt}}$ computed from the divergence, as given in column 10, appears to be in reasonable agreement with the more accurate 
result from the tracer particles. We discuss the issue of alternative measures of $P_{M,L}(\frac{ds}{dt})$ in more detail in the Materials and Methods section.
 
\subsection{Temporal Correlations}
\label{sec:tc}

Fig.\ \ref{fig:dsdt_correlation} shows the temporal correlation function of $\frac{ds}{dt}$ and its dependence on viscosity and Mach number.  The correlation function is defined as $\xi_{ds/dt} (\tau)\equiv \langle \frac{ds}{dt}(t)  \frac{ds}{dt}(t+\tau) \rangle$, which only depends on the time lag $\tau$ in a steady state.  The left panels of this figure adopt linear coordinates on the $y$-axis and logarithmic units on the $x$-axis, to emphasize 
the structure at small time scales. At the far left, $\xi_{ds/dt}(\tau)$ drops from  $\sigma_{{ds}/{dt}}^2$ at $\tau=0$ to negative values on a timescale much smaller than the eddy turnover time. 

Since $\sigma_{{ds}/{dt}}^2 = \xi_{ds/dt} (\tau =0)$ and the main contribution to $\sigma_{{ds}/{dt}}^2$ is due to shocks, the behavior of $\xi_{ds/dt}(\tau)$ for small $\tau$ is expected to also be shock dominated. Intuitively, such compressions  remain coherent only for the time it takes for a Lagrangian tracer to pass a shock width, which we label as $\Delta t_{\rm shock}$. The exact behavior of $\xi_{ds/dt}$ at $\tau \lesssim \Delta t_{\rm shock}$  depends on the average internal shock structure, and the negative dip at $\tau \approx \Delta t_{\rm shock}$  corresponds to an expansion, which may be viewed as a reaction to the strong compression the tracer just experienced. Finally, the anti-correlation at large time lag ($\xi_{ds/dt}(\tau)<0)$ is crucial to keep the variance of $s$ finite. 

\begin{figure*}[t]
\begin{center}
\includegraphics[width=1.0\textwidth]{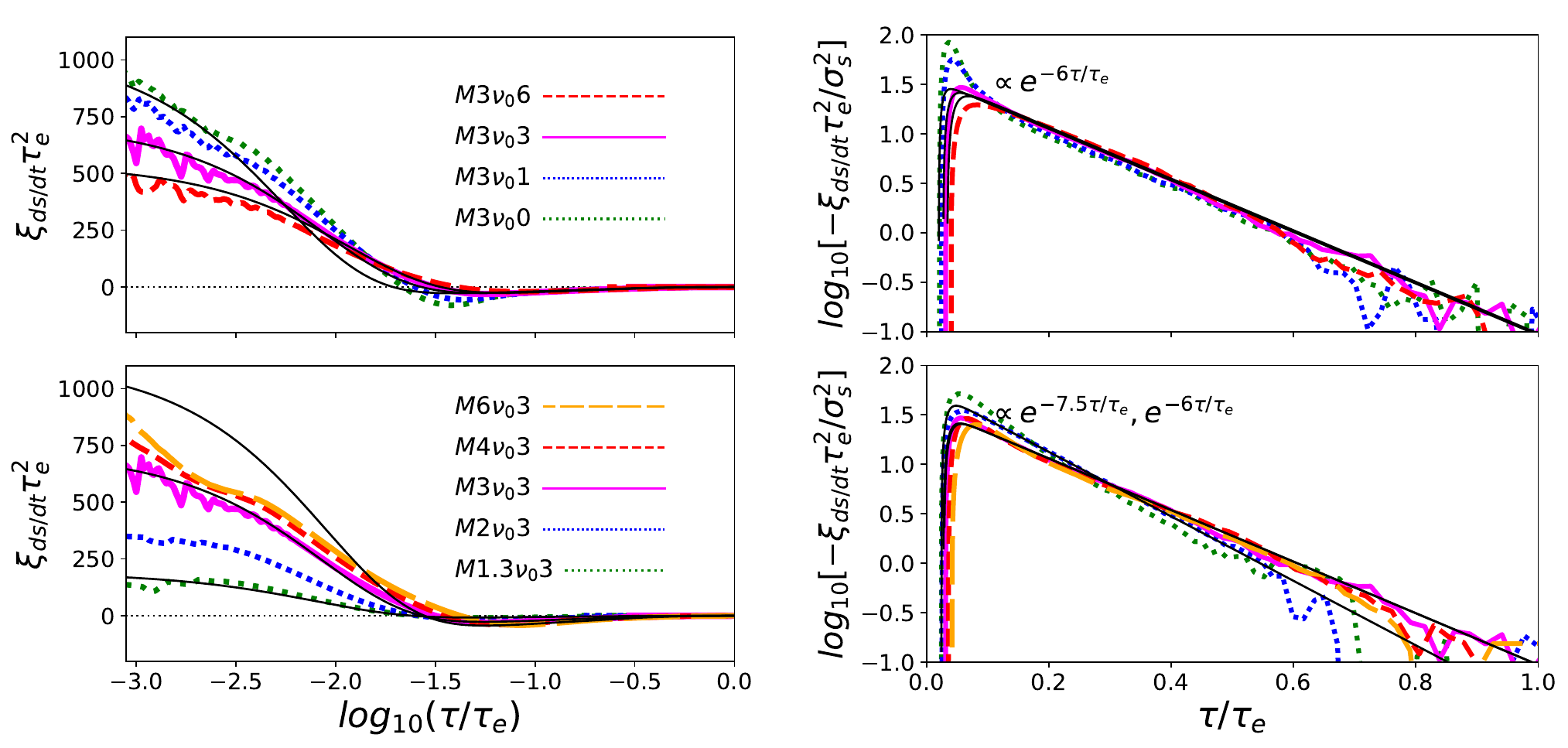}
\end{center}
\vspace{-0.2in}
\caption{{\bf Temporal correlation function of $\frac{ds}{dt}.$}  In the upper row, $\xi_{ds/dt}$ is
measured from the $M_{\rm rms} = 3.2$ simulations with different viscosities and in the lower row, $\xi_{ds/dt}$ is measured from $\nu_0 = 3\times10^{-4} L_{\rm box} c_s$
simulations with different Mach numbers. In each row, the panels on the left show the overall correlation function $\xi_{{ds}/{dt}} \times \tau_e^2$, while the panels on the \textbf{right} show the negative part of the correlation function, normalized by variance in $s,$ $\log_{10} [-\xi_{{ds}/{dt}} \times \tau_e^2/\sigma^2_s].$   The line styles and colors are labeled as in Fig.\ 2. The solid black lines show the results of fits to the stochastic differential model, eq.\ (\ref{eq:xidsdt}), with parameters as given in Table 
\ref{table:fits}.}
 \label{fig:dsdt_correlation}
\end{figure*}

At the lowest viscosities, $\Delta t_{\rm shock}$ is the shortest. Thus, in the upper panel, the M3$\nu_0$0 results drop most rapidly, while the M3$\nu_0$6 results show the most gradual decline.  In a realistic situation the shock width would only be several mean free paths, meaning that the variance of $ds/dt$ would be extremely large and $\Delta t$ shock would be extremely short. However, the integral of the short-time piece of the correlation function would be almost the same. On the other hand, increasing the Mach number while keeping the viscosity fixed increases the amplitude of the short-time correlation function, but has only a weak effect on $\Delta t_{\rm shock}$. 

As we derive in \S\ref{sec:analyticcorr},  $\int_{0}^{\infty} \xi_{{ds}/{dt}}(\tau)d\tau =0,$ meaning that the long-time integral of the negative part of the correlation function of $\frac{ds}{dt}$ must have the same amplitude as the short-time positive part. In \S\ref{sec:analyticcorr}, we also show that 
\be
\sigma_s^2= - \int_{0}^{\infty} \tau \xi_{{ds}/{dt}}(\tau) d\tau, 
\ee
which indicates the crucial role of the long-time behavior of the negative correlation function of $\frac{ds}{dt}$.  The equation shows that the dip in $\xi_{{ds}/{dt}}$  and its decay toward 0  at large time lag determine $\sigma_s^2$,  while the short time behavior at $\tau \lesssim \Delta t_{\rm shock}$ itself has no significant impact. 

The significant negative correlation at large time lag, which is crucial for determining the variance of $s$, also appears to be at odds with the assumption of independence  to explain the lognormal density distribution through the central limit theorem\cite{Passot98}. Intuitively, the density jumps cannot be completely independent, because the central limit theorem would otherwise predict an infinite density variance.   

\begin{table}
 \centering
\vspace{0.2in}
\hspace{-0.4in}
\begin{tabular}{|l|c|c|c|}
\hline
Name & $\alpha \tau_e$ & $\beta^2 \tau_e$ & $\Delta t_{\rm shock}/\tau_e$ \\
\hline
M3$\nu_0$0   & 6.0 & 11.5 & 0.0110   \\
M3$\nu_0$1   & 6.0 & 12.4 & 0.0130   \\
M3$\nu_0$3   & 6.0 & 13.1 & 0.0173   \\
M3$\nu_0$6   & 6.0 & 13.6 & 0.0236  \\
M1.3$\nu_0$3  & 7.5 & 3.24 & 0.0161   \\
M2$\nu_0$3   & 7.0 & 6.84 & 0.0158  \\
M4$\nu_0$3   & 6.0 & 17.0 & 0.0176   \\
M6$\nu_0$3   & 6.0 & 21.3 & 0.0181 \\
\hline 
\end{tabular}
\vspace{0.1in}
\caption{Fit parameters of our stochastic models with time-correlated noise.}
\label{table:fits}
\end{table}

In the right panels of Fig.\ \ref{fig:dsdt_correlation}  we show the long-time behavior of $-\xi_{ds/dt}$, now adopting a linear $x$-axis and logarithmic $y$-axis and normalizing each curve by $\sigma^2_s.$ Plotted in this way, all the curves exhibit an exponential behavior. Every curve is well fit by $-\sigma_s^2 \alpha^2 \exp(-\alpha \tau)$ where $\alpha$ is $6/\tau_e$ for all runs with Mach 3 or greater, and $\alpha$ is between $7.5/\tau_e$ and $7/\tau_e$ for runs with $M_{\rm rms} \leq 2.$ 

As we derive in  \S\ref{sec:langevin}, this exponential decline is a key feature of the Ornstein-Uhlenbeck (OU) process, which is a stochastic differential equation that defines how a system evolves when subjected to a stochastic forcing that is balanced by a linear, deterministic drift term.  In such a process, the fluctuations of $s$ evolve according to
\begin{equation}
ds = -\alpha (s - s_0) dt + \beta d W(t),
\label{eq:oumodel_s}
\end{equation}
where $W$ is the Wiener process, which for any $t \geq 0$ is made up of stationary, independent increments.In \S\ref{sec:discrete} we describe a closely-related stochastic process in which we replace white noise $dW(t)$ by noise with a finite correlation time, $\chi(t).$ We set the correlation function of $\chi$ to be exponential, i.e., $\langle \chi(t) \chi(t+\tau) \rangle = \frac{1}{\Delta t}\exp\left(-\frac{2|\tau|}{ \Delta t }\right)$ with a correlation time of $\Delta t/2,$ see also \cite{Sawford91}. As 
$\Delta t$ in the model controls how fast the correlation function $\xi_{ds/dt}$ drops to zero, we set $\Delta t = \Delta t_{\rm shock}$. 
The leads to
\begin{align}
\xi_{ds/dt}(\tau) = \frac{\beta^2}{2 [1-(\alpha \Delta t_{\rm shock}/2)^2]}\left[ \frac{2 \exp\left(-\frac{2 \tau}{\Delta t_{\rm shock}}\right) }{\Delta t_{\rm shock}} -\alpha \exp(-\alpha \tau) \right],
\label{eq:xidsdt}
\end{align}
which gives us a one-dimensional model to fit and interpret our simulation results. 
 As the equation predicts that $\xi_{ds/dt}(\tau)$ is exponential at large time lag, 
the exponent $\alpha$ may be measured by fitting the correlation function.  
Other model parameters can be calibrated using the relations to the variances of $s$ and $\frac{ds}{dt}$, 
\be
\frac{\beta^2}{1 + \alpha \Delta t_{\rm shock}/2}  = 2 \alpha \sigma^2_s = \Delta t_{\rm shock} \sigma^2_{ds/dt}.
\label{eq:fit}
\ee 
Note that, in the limit of small $\alpha \Delta t_{\rm shock},$ eq.\ (\ref{eq:fit}) is 
independent of the adopted time-correlated noise. Any form for the noise, whether based an exponential correlation (as above), 
a Gaussian correlation (as in eq.\ \ref{eq:descreteoudsdtcorrelation}), or any other correlation, will 
give $2 \alpha \sigma^2_s = \Delta t_{\rm shock} \sigma^2_{ds/dt} = \beta^2$ if $\alpha \Delta t_{\rm shock} \ll 1.$ 
The physical origins, properties, and impacts of these model parameters will be discussed in more detail in \S 3. 

In Table \ref{table:fits}, we give the results for measured parameters , quantifying the effect of Mach number and viscosity. 
As both $\alpha \tau_e$ and $\sigma_s^2$ are roughly constant across the runs with the same 
$M_{\rm rms},$ varying $\nu$ has only a small change on $\beta^2 \tau_e.$ However, viscosity has a direct impact on $\Delta t_{\rm shock}/\tau_e$, which more than doubles from run M3$\nu_0$0  to run M3$\nu_0$6. 
In all cases $0.03<\alpha \Delta t_{\rm shock}/2 < 0.1,$ justifying our use of a model with small but finite correlation time. 

On the other hand, $\Delta t_{\rm shock}/\tau_e$, varies $\lesssim 10\%$ between the M1.3$\nu_0$3 run and the M6$\nu_0$3 run.  Intuitively, $\Delta t_{\rm shock}$ is the time to cross a shock width, which is 
the same for the runs with the same viscosity, so that $\Delta t_{\rm shock}$ is proportional to the eddy crossing time $\tau_e$. 
Note that since both $\Delta t_{\rm shock,}/\tau_e$ and $\alpha \tau_e$ remain roughly 
constant at a fixed explicit viscosity, we expect from eq.\ (\ref{eq:fit}) that 
$\sigma_s^2$ and $\sigma_{ds/dt}^2$ would be roughly proportional to each other, as observed in Table \ref{tab:sigmas}. On the other hand, changing the Mach number has a strong impact on $\beta^2 \tau_e,$ which increases from 3.2 at $M_{\rm rms} =1.3$ to 21.3 at $M_{\rm rms} =5.9.$ Intuitively, this is because increasing $M_{\rm rms}$ boosts the amplitude of the stochastic forcing by shocks.

The black lines in Fig.\ \ref{fig:dsdt_correlation} compare the results of our simulation with the fits of the stochastic differential model for a range of simulations. At long times, the model with time-correlated noise provides an excellent description of all the data. At short times, the model also provides a good match to the data for runs with significant viscosities $\nu_0 \geq 3 \times 10^{-4} L_{\rm box} c_s$ and moderate Mach numbers $M_{\rm rms} \leq 3.$ However, for runs without explicit viscosity or when the Mach number is very high, the model matches $\xi_{ds/dt}$ at $\tau=0$ and the integral of $\xi_{ds/dt}$ over the range where $\xi_{ds/dt}>0,$ but it does not recover the measured shape of $\xi_{ds/dt}$ at short times. 

These discrepancies could be removed by adjusting the form of the time-correlated noise, but this would have only a minor effect on our fit parameters. Furthermore, the cases and the range of $\tau$ values in which the models are most discrepant are those in which shocks are most likely to be underresolved. Hence we do not attempt to adjust the correlation of time-correlated noise to account for these differences.  
The effect of numerical resolution is studied in more detail in \S \ref{sec:resolution}. 

Fig.\ \ref{fig:s_cor} shows the correlation function of $s$ measured from our simulations.  Recall that, by definition, this tracks the correlations of $s - \left<s \right>$ over time, subtracting the overall mean value. In the left panel of this figure, we compare this evolution with  
\be
\xi_s (\tau) = - \int\limits_0^\infty \tau'\xi_{ds/dt}(\tau+\tau')d\tau',
\label{eq:xiss}
\ee
as derived in \S \ref{sec:analyticcorr}. As our measurements of $\xi_{ds/dt}(\tau)$ are somewhat noisy at large values of $\tau/\tau_e,$ we replace $\xi_{ds/dt}(\tau)$ with the best fit exponential when computing the portion of the integral above $\tau = 0.5 \tau_e.$ This gives a good match in all cases, verifying eq. (\ref{eq:xiss}).

\begin{figure*}[t]
\begin{center}
\hspace{-0.05in}
\includegraphics[width=1.0\textwidth]{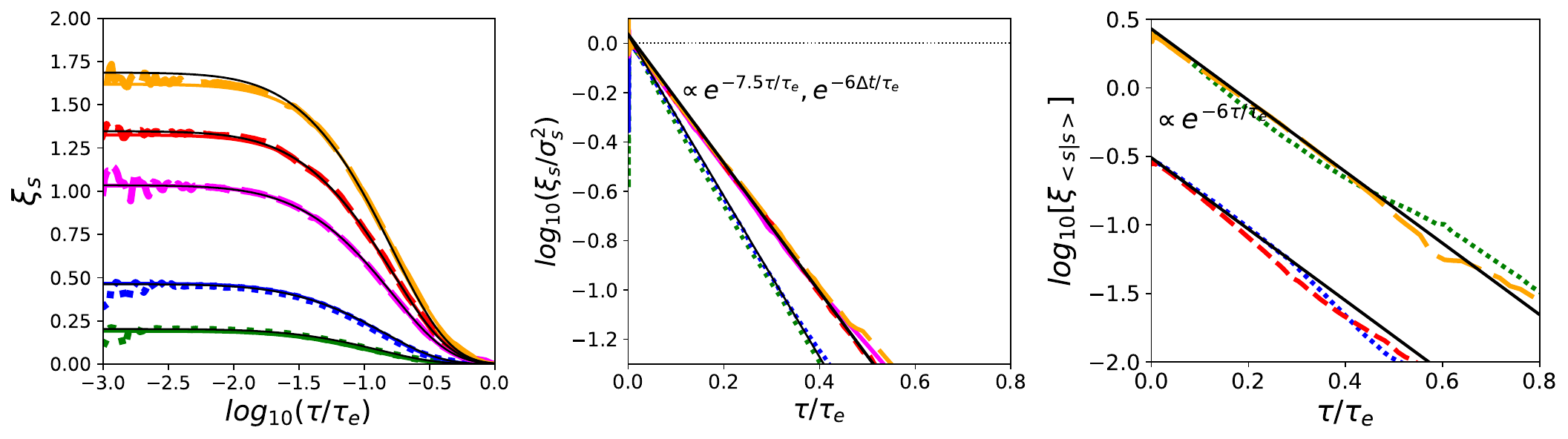}
\end{center}
\vspace{-0.2in}
\caption{ {\bf Correlation function of $s.$}   {\em Left:} Comparison between analytic predictions of $\xi_s(\tau/\tau_e)$ and results from  $\nu_0=3\times10^{-4} L_{\rm box} c_s$ simulations with different Mach numbers.
The simulation results are shown with colors and line styles as in Fig.\ \ref{fig:dsdt_correlation},  the colored solid lines show the results of the prediction from eq.\ (\ref{eq:xiss}), and the solid black lines give the predictions from the discretized random walk model.  {\em Center:} $\xi_s(\tau/\tau_e)$ normalized by $\sigma^2_s,$ from $\nu_0 = 3\times10^{-4} L_{\rm box} c_s$ simulations with different Mach numbers (colored lines), as compared to the discretized random walk model (solid black).  {\em Right:} The correlation function of particles from the M3$\nu_0$3 simulation split up into four bins according to their initial $s$ values.  Here the particles with $s \leq \left< s \right> - \sigma_s$, (green densely-dotted) at time $0,$ $\left< s \right> - \sigma_s \leq s \leq 0$, (blue dotted), $0 \leq s \leq \left< s \right> + \sigma_s$ (red short-dashed), $ \left< s \right> + \sigma_s \leq s$ (yellow long-dashed), are compared to the discretized random walk model (solid black).}
\label{fig:s_cor}
\end{figure*}

The central panel of Fig.\ \ref{fig:s_cor} emphasizes the long-time features of $\xi(\tau).$ Again, these results display an exponential decline in all cases, as required by the stochastic model. The black lines in the left and center panels of Fig.\ \ref{fig:s_cor} show the results of a comparison between our model with time-correlated noise, which provides a good fit to $\xi_s(\tau)$ at all times and for all simulation cases.

The short ($\tau_e/6$ to $\tau_e/7$) decorrelation times measured in our simulations are also consistent with the short times measured with the more indirect approach adopted in \cite{Scannapieco18} and \cite{Mocz19}. However, those studies found that a variable timescale was needed to explain the difference between the $s$ value of an Eulerian cell at a time $t$, and the $s$ value of the material from which it was previously comprised.

In the right panel of Fig.\ \ref{fig:s_cor}, we compare the behavior of the correlation functions of regions with various densities in our M3$\nu_0$3 simulation. Here we divide the particles into four groups by $s$ at $t=0$ and plot the forward correlation function describing their behavior at later times. These curves show that the decorrelation time is independent of the initial density and that the evolution of all four correlation functions is well described by an exponential decay with a decorrelation time of $\approx \tau_e/6.$ 

Note however that these functions describe very different behaviors. Although the exponential decline is the same for all of them, the magnitude of the changes is a strong function of $s.$ This means that, to maintain the overall evolution, the rate of change of $s$ must be faster in the tails of the distribution than near the mean. This is exactly the behavior expected in the stochastic model, as can be seen in eq.\ (\ref{ousolution}), which shows that any value $x$ reverts to mean according to an exponential with the same decorrelation timescale. We explore the impact of $s$ on the rate of change of $s$ in more detail below, showing that such conditional averages can be used to further elucidate the PDF of the density fluctuations.

\subsection{Conditional Averages}
\label{sec:conditional}

As described in more detail in  \S \ref{sec:analyticcond}, the conditional averages of $\frac{ds}{dt}$ and $\frac{d^2 s}{dt^2}$ as a function of $s$ provide us with an alternative approach to analyzing our simulations. The simplest such quantity is the average value of $\frac{ds}{dt}$ conditional on $s,$ $\left< \frac{ds}{dt} | s \right>,$ which is predicted to be exactly zero in a steady state, and thus serves as a test of the accuracy of our measurements \cite{Pan18}.   As shown in \S \ref{sec:accuracyconditional}, the departure of $\left< \frac{ds}{dt} | s \right>$ from zero, as measured by its ratio to $\sigma_{ds/dt},$ is much less than 1\% over the full region from $\left< s \right> -3.5 \sigma_s  \leq s \leq \left< s \right> + 2 \sigma_s$, and less than 5\% even in regions with $\left< s \right> + 2 \sigma_s \leq s \leq \left< s \right> + 3.5 \sigma_s$ in which there are significantly less particles.

\begin{figure}[t]
\begin{center}
\includegraphics[width=1.0\textwidth]{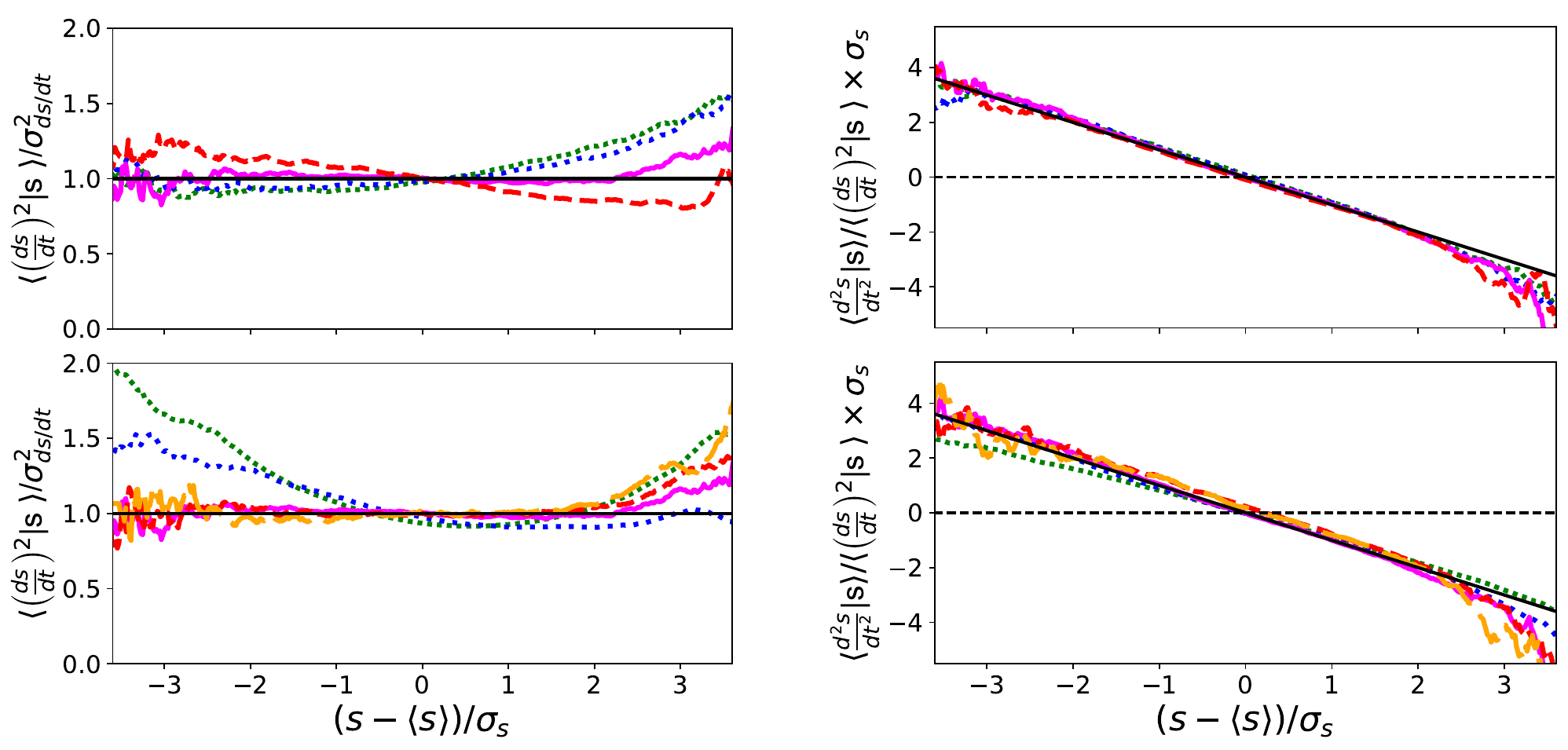}
\end{center}
\vspace{-0.2in}
\caption{{\bf Conditional averages as a function of $s.$} {\em Left:} The average value of $\left< (\frac{ds}{dt})^2 | s \right>$ as a function of  $(s-\left<s\right>)/\sigma_s,$  normalized by $\sigma_{{ds}/{dt}}^2.$ {\em Right:} The average value of the acceleration of the change in density $\left< \frac{d^2s}{dt^2} | s \right>$ as a function of $(s-\left<s\right>)/\sigma_s,$ normalized by $\left< (\frac{ds}{dt})^2 | s \right>/\sigma_s^2.$ The upper panels give results from runs with different viscosities and the lower panels give results with different Mach numbers. The line styles and colors are as in previous figures, while the black line gives the prediction from the stochastic model.}
\label{fig:DsdtsqandR}
\end{figure}

 As described in \S \ref{sec:analyticcond}, and confirmed in \S  \ref{sec:shape} the conditional averages of $(\frac{ds}{dt})^2$ and $\frac{d^2 s}{dt^2}$ determine the shape of the $P_{M}(s)$ as 
 \be
 P_{M}(s) \propto \exp \left[ \int_{-\infty}^s ds' \frac{\left< \frac{d^2 s}{dt^2} | s' \right>}{\left< \left(\frac{ds}{dt} \right)^2 | s' \right>} \right].
 \label{eq:shape}
 \ee
 A strength of the formulation based on conditional averages is that it is general and does not rely on the independence of density jumps and the central limit theorem as in \cite{Passot98} or the assumption of Gaussian noise as in our stochastic model.  
 
In the left panels of Fig.\ \ref{fig:DsdtsqandR}, we plot $\left< \left(\frac{ds}{dt}\right)^2 | s \right>.$ Like $\sigma^2_{ds/dt}$ and $\xi_{ds/dt},$ this term is dominated by rare events with large $\frac{ds}{dt}$ values, which correspond to shocks. 
The behavior of this conditional average is well described by the stochastic model, which predicts that $\left< \left(\frac{ds}{dt}\right)^2 | s \right> = \sigma^2_{ds/dt}$ for all values of $s$ (see eq.\ \ref{eq:OUdsdtsq}). Indeed, the changes in $\left< \left(\frac{ds}{dt} \right)^2 | s \right>$ are less than $\approx 10\%$ for most cases and over 
most $s$ values. However, for the cases with $M \geq 3,$ these differences increase somewhat at very high $s$ values, varying weakly with $\nu.$ Finally, the largest differences are seen for the transonic case at very low $s$ values, 
but even in this case, the deviations are less than a factor of 2 from the analytic predictions.  

In the right panels of  Fig.\ \ref{fig:DsdtsqandR} we show the behavior of $\left< \left(\frac{d^2 s}{dt^2}\right)| s \right>$ as a function of $\nu$ and $M,$ normalized by $\left< \left(\frac{ds}{dt}\right)^2 | s \right>.$
This plot shows that in all our simulations, shocks are systematically stronger in regions with densities below the mean density and weaker in regions with densities above the mean density. This process acts to narrow the $P_M(s)$ and sets its shape according to eq.\ (\ref{eq:shape}). A lognormal distribution of $s$ requires that the integrand in this equation behave as, 
\be
\frac { \left<\left(\frac{d^2s}{dt^2}\right)|s \right>}{\left<\left(\frac{ds}{dt}\right)^2|s \right>} = - \frac {\left(s-\left<s \right> \right)} {\sigma_s^2}. 
\label{eq:disdsdt2}
\ee
This provides a good fit for most cases and over most $s$ values (see the black solid line in right panels of Fig.\ \ref{fig:DsdtsqandR}), with one notable exception. At $s \gtrsim \left< s \right> + 3 \sigma_s,$  there is a significant downturn of $ \left< \frac{d^2s}{dt^2} | s \right>,$ making it clear that shocks are strongly weakened as they move into the densest regions. 

This means that the skewness in $P_{M}(s)$ arises from the rapid decrease in shock strength above $s - \left< s \right> \gtrsim 3 \sigma_s.$  Note that while this skewness is relatively modest for the mass-weighted PDF, it has a much greater impact on that volume-weighted PDF \cite{konstandin2012new}, especially at high Mach numbers. Since $P_V(s) \propto P_{M}(s) \exp(-s)$, when $\sigma_s$ is large, a shift towards lower $s/\sigma_s$ values leads to large changes in the volume fraction.

Inserting the results, Eq.\ (\ref{eq:fit}), for $\sigma_s$ and $\sigma_{ds/dt}$ from the correlated noise model in eq.\ (\ref{eq:disdsdt2}) yields, 
\be
\left<\left(\frac{d^2s}{dt^2}\right)|s \right> = - \left(s-\left<s \right> \right)\frac{2 \alpha}{\Delta t_{\rm shock}},
\label{eq:disdsdt3}
\ee
which connects the conditional acceleration of $s$ with the parameters $\alpha$ and $\Delta t_{\rm shock}$ of the correlated-noise model.  
Intuitively, a larger $\alpha$ corresponds to  a stronger dependence of shock acceleration with $s,$ which causes a faster reversion to the mean density. However, by taking the conditional average of the time-derivative of the Langevin equation,  we see that the contribution of the deterministic term to $\left<\left(\frac{d^2s}{dt^2}\right)|s \right>$ is given by $-\alpha \left<\frac{ds}{dt}|s \right> $, which is exactly zero at steady state.  This means that the effect of $-\alpha s$ term on the acceleration of $s$ enters instead indirectly through the noise term. 

From a different perspective, the detailed physical mechanisms behind the behavior of $\left<\left(\frac{d^2s}{dt^2}\right)|s \right>$ might be beyond the capability of the stochastic model to define. In reality, the $\frac{d^2s}{dt^2}$ is determined by shock acceleration into low-density regions and shock deceleration into dense regions \cite{Grover1966}. Modeling this processes in detail  requires understanding the interaction between forcing, shocks, and other density structures in the turbulent flow, which are not directly accounted for by the analytic model.

\section{Discussion}

While the current study cannot give a complete quantitative picture of the physical mechanisms that give rise to the parameters of the stochastic model, there are several important properties that provide new insights and suggestions for future investigations.

The first model parameter is $\alpha,$ which determines the average time that a particle in the flow ``forgets" its current density, reverting towards the mean. This contributes a rate of change of $s$ proportional to $- \alpha s$ and leads to exponential correlations for both $s$ and $\frac{ds}{dt}$ in excellent agreement with the simulation data. It is also notable that $\alpha$ is proportional to $1/\tau_e,$ the eddy turnover timescale, rather than $L_{\rm drive}/c_s,$ the timescale associated with changes in response to pressure fluctuations. This means that this process is associated with the shocks and expansions that produced inherently by the nonlinear interactions in turbulence, rather than simply with the regulation of density fluctuations by the pressure differences. Because $\alpha$ corresponds to a timescale that is $6-7$ times shorter than $\tau_e,$ the timescale for the flow to revert to the mean is relatively short, meaning that this process acts more rapidly than often assumed in the literature \cite{Krumholz05, Padoan11, Hennebelle11, Federrath12}. 

Finally, from our conditional averages, we find that $\alpha$ can also be thought of as a measure of shocks accelerating into low-density regions and decelerating as they move into high-density regions.  When viewed in this way, $\alpha$ is again extremely regular, apart from the downturn at $s-\left<s \right> \gtrsim 3 \sigma_s.$ This indicates a strong decrease in shock strength above this value, likely because, at these high densities, the pressure is strong enough to slow shocks to subsonic speeds. This feature causes the skewness in $P_M(s)$

\begin{figure}
\begin{center}
\includegraphics[width=0.48\textwidth]{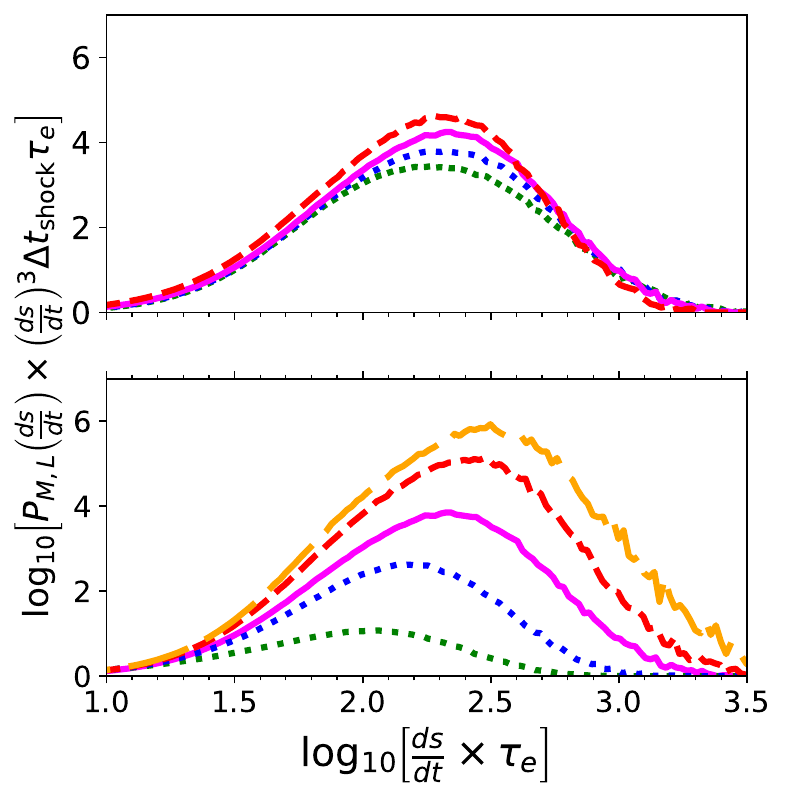}
\end{center}
\vspace{-0.15in}
\caption{{\bf The contribution to the variance of $s$ resulting from different values of $\frac{ds}{dt}.$} The top panel shows  measurements of  $P_{M,L}\left(\frac{ds}{dt} \right) \times \left(\frac{ds}{dt} \right)^3 \Delta t_{\rm shock}$ for Mach 3 runs with different viscosities, and the bottom panel shows the results of $\nu_0 = 3 \times 10^{-4} L_{\rm box} c_s$ runs with different Mach numbers. Colors and line styles are as in previous plots.}
\label{fig:betabreakdown}
\end{figure}

The second parameter in our model is $\beta^2,$ which sets the degree of stochastic driving. Because $\alpha \tau_e$ is nearly constant, especially at $M_{\rm rms} \geq 2,$ it is largely this parameter that sets 
the dependence of $\sigma_s^2$ on Mach number. 
In \S \ref{sec:pdf} we saw that $\sigma_{ds/dt}^2$ is almost completely set by compressions. This means that the compressive part of the probability distribution of $\frac{ds}{dt}$ provides us with information about which shocks are most important in determining $\sigma_s^2.$ In Fig.\ \ref{fig:betabreakdown} we show this distribution as the integrand of
\be
\frac{\beta^2}{1+\alpha \Delta t_{\rm shock}/2} = \int_0^{\infty} d \ln \left(\frac{ds}{dt} \right)
P_{M,L}\left(\frac{ds}{dt} \right) \times \left(\frac{ds}{dt} \right)^3 \Delta t_{\rm shock}.
\ee
The upper panel shows that changing $\nu$ has only a minor impact on this distribution. On the other hand, the lower panel of this figure shows a clear systematic trend between the simulations with different Mach numbers. 

As $M_{\rm rms}$ increases, the peak and cutoff in the distribution move to the right. The shift in $\frac{ds}{dt} \times \tau$ is $\appropto \Delta s = \ln(M^2)$ the change in $s$ associated with a shock of Mach number $M$.  Because the $P_{M,L}(ds/dt) \appropto \left(\frac{ds}{dt} \right)^{-3}$ at large $\frac{ds}{dt}$ values for the runs with $M \gtrsim 2,$ the area under this curve (and hence $\beta^2$) also increases approximately linearly with $\ln (M^2).$ At the lowest Mach number, the overall distribution of compressions drops significantly even at small $\frac{ds}{dt}$ values. This is likely due to the case that the shocks in the flow have a range of velocities and that many of these will drop below $c_s$, meaning that they will not drive significant compressions. 

The final parameter in our stochastic differential models is the correlation timescale of the noise, which we associate with the timescale of a typical compression, $\Delta t_{\rm shock}$. Intuitively,  $\Delta t_{\rm shock}$ corresponds to the time to pass the thickness of a shock. This association gives a good match between the model prediction and simulation result  for the shape of $\xi_{ds/dt}(\tau)$ at all times for cases in which viscosity is large and the Mach number is $\lesssim 3.$ For large Mach numbers and low viscosities, the match between the small time behavior of the model and the simulations is less close, although these are the cases in which shocks are most likely to be underresolved, and further simulations are needed to determine if this discrepancy is physical. If it is, it could be removed by adjusting the color of the noise used to drive the random walk, although this would have only a minor impact on our measurement of $\Delta t_{\rm shock}$ and no direct impact on the overall evolution of the density fluctuations.

Our results are limited to the fundamental case of solenoidally-driven, isothermal turbulence, but they provide a natural launching point to incorporate additional effects such as magnetic fields, which will alter the relationship between $M_{\rm rms}$ and density, and compressive driving, which will lead to an additional term associated with expansions/compressions caused directly by the driving modes \cite{Pan19}. Recently \cite{Appel23} addressed some of these issues by measuring the divergence of the velocity field to constrain the distribution of $\frac{ds}{dt}$ as a function $s$ in simulations of giant molecular clouds that included gravity, turbulence, stellar feedback, and magnetic fields. Our results suggest that simulations with tracer particles can be used to develop a better understanding of the impact of these processes.

Finally, by extending the model to the two-dimensional phase-space of temperature and density, a similar set of tracer particle measurements can likely be combined with a two-dimensional stochastic model to study the evolution of non-isothermal turbulence
\cite{Federrath15,Nolan15,Gray15,Buie18}. This has the potential to address such widespread issues as combustion in scramjets \cite{Chang18}, the structure of explosive volcanic eruptions \cite{Ogeden08}, the evolution of core-collapse supernovae \cite{Moesta15}, molecule formation in molecular clouds \cite{Glover10,Walch15}, the interstellar circulation of material \cite{Kim17}, and the origin of nonequilibrium ionization features in the circumgalactic medium \cite{Richings14, Buie18, Buie20}. The method developed here may also be applied to study the non-lognormal 
density PDF in non-isothermal turbulence \cite{Passot98} and in compressively driven turbulence \cite{Federrath10}. Future efforts will be devoted to developing stochastic differential equation models that are capable of handling these issues.

\section{Materials and Methods}

\subsection{Simulations}
\label{sec:sims}

\subsubsection{Turbulent Driving and Hydrodynamic Solver}
\label{sec:turbulentdriving}

The continuity and momentum equations evolved in our simulations are  
 \begin{equation}
 \frac{\partial \rho}{\partial t} +  \frac{\partial \rho v_i}{\partial x_i}  = 0,
\label{eq:continuity}
 \end{equation} 
and 
\begin{equation}
\frac{\partial v_i }{\partial t} +  v_j \frac{\partial v_i }{\partial x_j}= - \frac{1}{\rho} \frac{\partial p}{\partial x_i} + \frac{1}{\rho} 
\frac{\partial \sigma_{ij}}{\partial x_j} + a_{OU,i} (\bs{x}, t),
\label{eq:velocity}
\end{equation} 
where $p(\bs{x}, t)$ is the pressure, $\sigma_{ij}$ is the viscous stress tensor, and $ \bs{a_{OU}}(\bs{x}, t)$ is the driving accerleration. 
For an ideal gas, the bulk viscosity is $\sigma_{ij} = \frac{1}{2} \rho \nu ( \partial_i v_j +\partial_j v_i - \frac{2}{3} \partial_k v_k \delta_{ij}) $
where $\nu$ is the kinematic viscosity.

We drove the flow with a solenoidal acceleration field (i.e.\ $\partial_i {a_{i} } = 0$), that we imposed on the box at each time step \cite{Eswaran88, Benzi08}. We selected a range of wave numbers for driving, and for each Fourier mode within this range, we selected four numbers from independent Ornstein-Uhlenbeck (OU) processes,  each describing the real and imaginary parts of the acceleration in the two directions perpendicular to the wavevector. 
Our driving scheme can be summarized as $\left< a_i (k,t) a_j (k,t') \right>_E = P_a (k)(\delta_{i,j} - k_i k_j/k^2 ) \exp(- |t-t'|/\tau_f),$ where we chose the magnitude to be constant in the range of wave numbers 1 $\le L_{\rm box} |k|/2 \pi \le$ 3, such that the average forcing wavenumber was $k_f \simeq 4 \pi/L_{\rm box}.$ 
In other words, we chose the range of wavenumbers such that the driving scale of the turbulence was $\ell \simeq 2 \pi / k_f = L_{\rm box}/2.$ 
In all our simulations, we set the correlation time $\tau_f$ to 0.2 sound crossing times as we have in previous studies \cite{Pan10, Pan11, Pan12, Scannapieco18}. Note, however, that in some literature studies, the acceleration of each mode is varied independently at every time step \cite{Lemaster09}.

Following our previous turbulence studies \cite{Pan10, Gray15, Gray16, Buie20} we solved the hydrodynamic equations with the FLASH code \cite{Fryxell00}, version 4.2.5. We used an unsplit solver with third-order reconstruction \cite{Lee13} and employed a hybrid Riemann solver, which uses both an extremely accurate but somewhat fragile Harten-Lax-van Leer-Contact (HLLC) solver \cite{Toro94} and a more robust, but more diffusive Harten Lax and van Leer (HLL) solver \cite{Einfeldt91}. In all simulations, we specified a nearly isothermal equation of state, with $\gamma=1.0001,$ such that fixing the driving amplitude led to a roughly constant Mach number. In addition, the runs made use of an explicit kinematic viscosity, so that we were able to study the impact of varying this value.

\subsubsection{Effective Viscosity}
\label{sec:effectivenu}

To compute the effective viscosity of our simulations, we followed the approach described in \cite{Pan10} and considered the equation for the average kinetic energy per unit volume
\begin{eqnarray}
\frac{\partial}{\partial t} \langle \frac{1}{2} \rho v^2 \rangle_E
& =&  \left \langle p \frac{\partial v_i}{\partial x_i} \right \rangle_E  + \langle \rho f_i v_i \rangle_E \nonumber \\
& & - \frac{1}{2} \left \langle \rho \nu \left( \frac{\partial v_i}{\partial x_j} + \frac{\partial v_j}{\partial x_i}- 
\frac{2}{3} \frac{\partial v_k}{\partial x_k}\delta_{ij} \right)^2  \right \rangle_E,
\label{eq:keperunitmass}
\end{eqnarray}
which is derived from the momentum and continuity equations.
Here $\tilde{\rho}$ is the ratio of the local mass density to the average mass density and the ensemble
average in the Eulerian frame, $\left< ... \right>_E,$ is equal to the average over the flow domain for statistically homogeneous flows, as in our simulations. The three terms on the right-hand side of eq.\ (\ref{eq:keperunitmass}) represent the $pdV$ work, the energy injection from the driving force, and the viscous dissipation of kinetic energy, respectively. Note that eq.\ (\ref{eq:keperunitmass}) would include an additional term in magnetized turbulence, as the energy dynamics in that case are more complex than in the case we are considering here \cite{Grete23}.

\cite{Pan19} proved that for an isothermal turbulent flow, or more generally a barotropic flow, the average rate of the $pdV$ work, which is a reversible conversion between kinetic and thermal energies, is expected to be zero in a steady state. This is indeed confirmed by simulations with a sufficiently high viscosity that it can stabilize all shocks by itself \cite{Pan19}. However, for simulations using a Riemann solver without explicit viscosity, the average $pdV$ work computed from simulation data appears to be negative and significant especially at high Mach numbers, incorrectly suggesting that the $pdV$ preferentially converts kinetic energy to thermal energy. \cite{Pan19} pointed out that this wrong impression was because, to numerically stabilize shocks, the Riemann solver must artificially modify the flow quantities at the cell scale around strong discontinuities. A consequence is that the spatial derivatives of the flow quantities, such as the divergence, cannot be accurately computed from the simulation data, even though the Riemann solver conserves mass, momentum, and energy at high accuracy. 

A significant negative average $pdV$ work may only indicate the inaccuracy of spatial gradients in the simulation data; it does not necessarily suggest that the Riemann solver converts kinetic energy to thermal energy through $pdV$ work \cite{Pan19}. Therefore, the ratio of the two terms, $ \left \langle p \partial_i v_i \right \rangle_E$ and $\langle \rho f_i v_i \rangle_E$, in eq.\ (\ref{eq:keperunitmass}) may be used as an indicator for the accuracy of the measured divergence from the simulation data, and this quantity is given in Table \ref{tab:pdV} for each of our simulations. This ratio does drop by more than a factor of 2 between our M3$\nu_0$0 simulation and the M3$\nu_0$6 run with the largest explicit viscosity. However, it remains significant even in the most extreme case, suggesting that the solver still needs to artificially stabilize shocks. This indicates that shocks cannot be considered to be fully resolved in this study.

\begin{table*}[t]
 \centering
\begin{tabular}{|l|c|c|c|c|c|c|c|c|c|}
\hline
\, \, Name & $-\left \langle p \partial_i v_i \right \rangle_E $\\
\, \,           & $\left( \langle \rho f_i v_i \rangle_E \right)$ \\
\hline
M3$\nu_0$0   & 0.41  \\
M3$\nu_0$1   & 0.32  \\
M3$\nu_0$3   & 0.21  \\
M3$\nu_0$6   & 0.16 \\
M1.3$\nu_0$3  & 0.17  \\
M2$\nu_0$3   & 0.19  \\
M4$\nu_0$3   & 0.23  \\
M6$\nu_0$3   & 0.21 \\
\hline 
\end{tabular}
\vspace{0.1in}
\caption{The ratio of the average $pdV$ work term to the average energy input in our simulations.}
\label{tab:pdV}
\end{table*}%

It is of interest to estimate the effective total (numerical + explicit) viscosity in our simulations, even though the velocity gradient cannot be measured at high accuracy. In a steady state, we can set the time derivative term in eq.\ (\ref{eq:keperunitmass}) to zero, and considering that the $pdV$ work term is expected to be zero theoretically, we solve for the value of effective $\nu$ by balancing the viscous dissipation rate and the energy input rate, $\langle \rho f_i v_i \rangle_E$, in eq.\ (\ref{eq:keperunitmass}).  For each of our runs, we calculate the effective viscosity from 10 files, spaced equally throughout the steady state phase. These values are given in Table \ref{tab:runs} and referred to throughout this study.

\subsection{Analytic Framework}

In this section, we present four types of analytic results that we use extensively in interpreting our simulations. In \S \ref{sec:analyticcorr} we present new analytic results that relate the correlation functions of any quantity $x$ and its time derivative $\frac{dx}{dt}$; in \S \ref{sec:analyticcond} we summarize literature results regarding the conditional averages of the time derivatives of $s;$ and in \S \ref{sec:langevin} and \S \ref{sec:discrete} we present analytic predictions for the correlation functions and conditional averages for simple continuous and time-correlated stochastic models. 

\subsubsection{Correlation Functions}
\label{sec:analyticcorr}

In this subsection, we derive analytic results that relate the correlation functions of any quantity $x$ and its time derivative, $\frac{dx}{dt}$ in a steady state. These have not appeared in the literature before to our knowledge.  To derive them, we first define the correlation of $x$ at two times $t_1$ and $t_2$ as $\xi_x \equiv \langle (x(t_1)-\bar x) (x(t_2) -\bar x)\rangle$, where $\bar x$ is the mean of $x$, which is constant in a steady state. Similarly, we define the correlation function of $\frac{dx}{dt}$ as $\xi_{dx/dt} \equiv \langle \frac{dx}{dt}(t_1) \frac{dx}{dt}(t_2) \rangle$, such that $\frac {d^2 \xi_x}{dt_1 dt_2} = \xi_{dx/dt}$.

In a steady state, $\xi_x$ is a function of the time lag $\tau = t_2 -t_1$ only, i.e., it depends on $t_2$ and $t_1$ only through their difference $\tau$. As the same applies to $\xi_{dx/dt},$ it immediately follows that
\be
\frac {d^2 \xi_x (\tau)}{d\tau^2} = -\xi_{dx/dt}(\tau).
\ee
Integrating this equation once and considering that $\frac {d \xi_x (\tau)}{d\tau} \to 0$ as $\tau\to \infty$, we have
\be
\frac {d \xi_x (\tau)}{d\tau} = \int\limits_\tau^\infty \xi_{dx/dt}(\tau')d\tau'.
\label{eq:intermediate}
\ee
It is interesting to note that at zero time lag, $\frac {d \xi_x (0)}{d\tau} 
=\frac{1}{2} \frac{d\langle (x(t_1)-\bar x) ^2 \rangle}{dt_1} $, which vanishes in a steady state. Setting $\tau$ in eq.\ (\ref{eq:intermediate}) to zero then indicates that 
\be
\int\limits_0^\infty \xi_{dx/dt}(\tau')d\tau'=0, 
\label{eq:xiintis0}
\ee
suggesting that $\xi_{ds/dt}$ must become negative at large time lags. Integrating eq.\ (\ref{eq:intermediate}) again gives, 
\be
 \xi_x (\tau) = - \int\limits_\tau^\infty d\tau'\int\limits_{\tau'}^\infty \xi_{dx/dt}(\tau'')d\tau'',
\label{eq:intermediate2}
\ee
where we used the boundary condition that $\xi_x (\tau) \to 0$ as $\tau\to \infty$. 

Integrating by parts, we arrive at the final result 
\be
\xi_x (\tau) = - \int\limits_0^\infty \tau'\xi_{ds/dt}(\tau+\tau')d\tau',
\label{eq:xis}
\ee
where we have implicitly assumed that $\xi_{dx/dt}(\tau)$ approaches zero rapidly as $\tau\to \infty$. Setting $\tau$ to zero in this expression gives the variance of $x$, 
\be
 \sigma_x^2 = - \int_{0}^{\infty} \tau \xi_{dx/dt}(\tau) d\tau. 
 \label{eq:sigmas}
\ee
This equation shows that it is not the short time behavior of $\xi_{dx/dt}$ that determines the variance of $x$, but rather the much longer time lags when $\xi _{{dx}/{dt}}$ reverts from negative to 0. 
We give a more detailed discussion of this relation in \S \ref{sec:tc} where we use eqs.\ (\ref{eq:xis}) and (\ref{eq:sigmas}) to better understand our simulation results.

\subsubsection{Conditional Averages}

\label{sec:analyticcond}

In this subsection, we consider statistical quantities that provide insight into the shape of the probability distribution function of $s$, $P_{M,L}(s).$  These are complementary to the correlation function of $\frac{ds}{dt}$ discussed above, which captures the temporal evolution of $s$, but only constrains the variance of $P_{M,L}(s)$. 

Studies have shown that in a steady state $P_{M,L}(s)$ may be related to the ensemble average of the time derivatives of $s$ conditioned on the value of $s$ \cite{Pope93}. 
While previous studies examined the conditional averages in the Eulerian frame \cite{Pan18}, here we take the Lagrangian approach, although, in the case of conditional averages, the results in the Lagrangian and Eulerian frames are equivalent.

In a steady state, $P_{M,L}(s)$ is time-invariant, i.e., $d P_{M,L}(s)/dt =0$, which requires that the net probability flux into and out of each $s$ bin to be zero \cite{Pan18}. Intuitively, the probability flux is given by $P_{M,L}(s) \left< \frac{ds}{dt} | s \right> $, where $\left< \frac{ds}{dt} | s \right>$ is the average value of $\frac{ds}{dt}$ of the Lagrangian particles within a given $s$ bin. Setting this flux to zero in a statistical steady state leads to
\be
\left< \frac{ds}{dt} | s \right> = 0,
\label{eq:dsdtzero}
\ee
for all $s$. This is equivalent to the result of \cite{Pan18} that the conditional average of the velocity divergence, $\langle \nabla\cdot \bs v |s\rangle =0$ at all $s$.

To maintain a steady state, the time derivative of the probability flux must also be zero, which guarantees that $\frac{d^2 P_{M,L}(s)}{dt^2} =0$. 
This results in a second condition 
\be
 P_{M,L}(s) \left< \frac{d^2 s}{dt^2} | s \right> 
 -\frac{d}{ds} \left[ P_{M,L}(s) \left< \left(\frac{ds}{dt} \right)^2 | s \right>  \right] = 0. 
\label{eq:second}
\ee
Here, the time derivative of flux
consists of two contributions, one from mechanisms that
change $\frac{ds}{dt}$ and the other from mechanisms that change $s.$  Mechanisms that change $\frac{ds}{dt}$ lead to the first term in eq.\ (\ref{eq:second}) in a straightforward way. Mechanisms that change $s$ affect the ``local'' distribution of $s$ at a rate proportional to $\frac{ds}{dt}$, 
 and because there is already a factor of $\frac{ds}{dt}$ in the flux, this gives rise to the second
 term in eq.\ (\ref{eq:second}) which is proportional to $\langle(\frac{ds}{dt})^2|s\rangle$.  
In a steady state, the two contributions must balance each other, leading to eq.\ (\ref{eq:second})
\cite{Pope93,Pan19}. This can be solved to give
\be
P_{M,L}(s) \propto \frac{1}{\left< \left(\frac{ds}{dt} \right)^2 | s \right>} \exp \left[ \int_0^s ds' \frac{\left< \frac{d^2 s}{dt^2} | s' \right>}{\left< \left(\frac{ds}{dt} \right)^2 | s' \right>} \right],
\label{eq:psconditional}
\ee
which we use in interpreting our simulation results.

\subsubsection{The Langevin Model}

\label{sec:langevin}

Next, we consider the main features of a simple Langevin model, which is a stochastic differential equation that defines how a system evolves when subjected to a combination of deterministic and fluctuating forces. The most famous application of this model is to describe the Brownian motion of a particle suspended in a medium due to random collisions \cite{Lemons97}, and it has been applied in many other fields including electrical engineering biology, chemistry, and finance \cite{Johnson28, Hofling13, Gillespie01, Vasicek77}.  

The model forms the basis of the driving scheme used in many direct numerical simulations of turbulence \cite{Eswaran88, Benzi08}, and it has been applied in limited cases to study the temporal correlations of turbulent velocity in the Lagrangian frame \cite{Novikov89, Pedrizzetti94, Pedrizzetti99, Renner01}. It has also been adopted by \cite{Scannapieco18} and \cite{Mocz19} to investigate density fluctuations in compressible turbulence. 

Here we summarize several key properties of the Langevin model, limiting ourselves to one of the simplest and most widely applied versions of the model, the Ornstein-Uhlenbeck (OU) process. In this case, the fluctuations of a quantity $x$ are modeled as
\begin{equation}
dx = -\alpha (x - x^*) dt + \beta d W(t).
\label{eq:oumodel}
\end{equation}
Here, $x^*$ is the steady-state mean value of $x,$ and $W$ is the Wiener process, which for any $t \geq 0$ is made up of stationary, independent increments, and any increment $W(t+\Delta t)-W(t)$ is a Gaussian with a standard deviation of $\Delta t.$ To simplify the analysis, we consider a case in which $ x^*=0,$ but the resulting expressions can also be applied to a system in which this is not the case by simply replacing $x$ with $x-x^*.$

With $ x^*=0,$ eq.\ (\ref{eq:oumodel}) can be solved as
\begin{equation}
x(t) = x(0) \exp(-\alpha t) + \exp(-\alpha t) \beta \int\limits_0^t \exp(\alpha t') dW(t'). 
\label{ousolution}
\end{equation}
At late times, the first term becomes negligible, and the second term indicates a Gaussian distribution for $x$.
 Ignoring the first term in eq.\ (\ref{ousolution}), we calculate the temporal correlation function of $x$ as 
\begin{align}
\xi_x(\tau) \equiv \langle x(t+\tau)&x(t) \rangle = \exp(-2\alpha t-\alpha \tau) \beta^2 \times \notag \\
& \left\langle \left( \int\limits_0^{\min(t, t+\tau)} \exp(\alpha t_1) dW(t_1) \right)^2 \right\rangle, 
\end{align}
where we have made use of the property of the Wiener process that $\langle (W(t') -W(t)) W(t) \rangle =0$ for all $t'\ge t.$
Using Ito isometry \cite{vanKampen92,Gardiner94}, the above equation can be rewritten as
\begin{equation}
\xi_x(\tau) = \exp(-2\alpha t-\alpha \tau) \beta^2 
\int\limits_0^{\min(t, t+\tau)} \exp(2\alpha t_1) dt_1, 
\end{equation}
which can be integrated to give 
\begin{align}
\xi_x(\tau) = \frac{\beta^2}{2\alpha}\ \exp(-\alpha |\tau|),
\label{oucorrelation}
\end{align}
at large times. Thus, when a steady state is reached, the variance of $x$ is $\sigma_x^2 = \xi_x(0) = 
 \frac{\beta^2}{2\alpha}$, and the time correlation is an exponential with a characteristic decorrelation time of $\alpha^{-1}.$

In a steady state, the correlation functions of $x$ and its time-derivative, $\frac{dx}{dt}$, are related by 
$\xi_{dx/dt}(\tau) \equiv \langle \frac{dx}{dt} (t+\tau) \frac{dx}{dt}(t) \rangle = - \frac{d^2}{d \tau^2 } \xi_{dx/dt}(\tau) $. Taking the second derivative of eq.\ (\ref{oucorrelation}) with respect to $\tau$ yields, 
\be
\xi_{dx/dt}(\tau) = \sigma_x^2
\left[ 2 \alpha \delta (\tau) - \alpha^2 \exp(-\alpha |\tau|) \right], 
\label{oudsdtcorrelation}
\ee
where $\delta$ denotes the Dirac delta function, which originates from the fact that the first derivative of $\exp(-\alpha |\tau|) $ is discontinuous at $\tau=0$. It is also straightforward to verify that the Langevin model satisfies that
$\int_0^\infty \xi_{dx/dt}(\tau) d \tau =0$ and $\sigma_x^2 = - \int_0^\infty \tau  \xi_{dx/dt}(\tau) d \tau $, as required by eqs.\ (\ref{eq:xiintis0}) and (\ref{eq:sigmas}) above.

\subsubsection{Langevin Model with Time-Correlated Noise}

\label{sec:discrete}

One issue with the Langevin model is that setting $\tau$ to zero in eq.\ (\ref{oudsdtcorrelation}) gives an infinite variance for $dx/dt$. This is because the path of a Wiener process is non-smooth. In other words, the random driving source for $x$ is white noise, since $dW$ can be written as $dW = \xi dt$, where $\xi$ is white noise. To remedy this non-physical problem, one may replace the white noise with a discretized random walk, where each step lasts a duration $\Delta t.$ In this case, the stochastic term becomes time-correlated noise.

If the force is drawn from a Gaussian distribution with an amplitude of $\beta/\sqrt{\Delta t}$ this gives
\begin{align}
\xi_{dx/dt}(\tau) = & \frac{\beta^2}{2 \alpha \Delta t^2}[2 \exp(-\alpha \tau) - \exp(-\alpha \Delta t +\alpha \tau)  - \exp(-\alpha \Delta t - \alpha \tau)] \hspace {.3 cm }{\rm for} \hspace {.2 cm } \tau \le \Delta t  \notag\\
 & \frac{\beta^2}{2 \alpha \Delta t^2}[ 2 \exp(-\alpha \tau) - \exp(\alpha \Delta t -\alpha \tau) -\exp(-\alpha \Delta t - \alpha \tau)] \hspace {.3 cm } {\rm for} \hspace {.2 cm } \tau > \Delta t.
\label{eq:descreteoudsdtcorrelation}
\end{align}
It is straightforward to show that eq.\ (\ref{eq:descreteoudsdtcorrelation}) approaches eq.\ (\ref{oudsdtcorrelation}) in the limit $\Delta t \to 0.$ However, unlike eq.\ (\ref{oudsdtcorrelation}), 
eq.\ (\ref{eq:descreteoudsdtcorrelation}) gives a finite variance for $\frac{dx}{dt}$, 
$\langle (\frac {dx}{dt})^2 \rangle = \frac{\beta^2}{\alpha \Delta t^2}[1-\exp(-\alpha \Delta t)],$
which approaches $\frac {\beta^2}{\Delta t}$ in the limit of $\alpha \Delta t \ll 1.$

To solve the problem of infinite variance for $\frac{dx}{dt}$, in principle, we may consider a model with any time-correlated noise. Above we show that our simulation results are well approximated by a case in which we set the correlation function of $\chi$ to  $\langle \chi(t) \chi(t+\tau) \rangle = \frac{1}{\Delta t}\exp\left(-\frac{2|\tau|}{ \Delta t }\right)$ with a correlation time of $\Delta t/2.$ This leads to
\begin{align}
\xi_{dx/dt}(\tau) = \frac{\beta^2}{2 [1-(\alpha \Delta t/2)^2]}\left[ \frac{2}{\Delta t} \exp\left(-\frac{2 \tau}{\Delta t}\right) -\alpha \exp(-\alpha \tau) \right],
\label{eq:colordsdtcorrelation}
\end{align}
which again gives a finite variance for $\frac{dx}{dt}$, $\langle (\frac {dx}{dt})^2 \rangle = \frac {\beta^2}{\Delta t (1+\alpha \Delta t/2)}$. In this case, the corresponding variance of $x$ is $\sigma_x^2 = \frac{\beta^2}{2 \alpha (1+\alpha \Delta t/2)}.$
In the limit of $\alpha \Delta t \ll 1,$ $\sigma_x^2$ and $\langle (\frac {dx}{dt})^2 \rangle$ approach $\frac{\beta^2}{2\alpha}$ and
$\frac{\beta^2}{\Delta t},$ respectively. In this model, the noise $\chi$ has a variance of $1/{\Delta t}$, and thus the amplitude of each step in the random walk for $x$ is $\beta/\sqrt{\Delta t}$.
We point out that the bi-exponential form as in eq.\ (\ref{eq:colordsdtcorrelation}) was previously proposed to model the Lagrangian velocity correlation in incompressible turbulence \cite{Sawford91}.

For the OU process, we can also derive several useful results for the conditional averages of $\frac{dx}{dt}$ and $\frac{d^2 x}{dt^2},$ which are of interest for understanding the shape of the probability distribution function, as discussed in \S \ref{sec:analyticcond}.
For both the Langevin model and the time-correlated version, $x$, $\frac{dx}{dt}$ and $\frac{d^2x}{dt^2}$ are all Gaussian. In a steady state, we have $\langle x \frac{dx}{dt} \rangle = \frac{1}{2}\frac{\langle x^2 \rangle}{dt} =0$, which means $x$ and $\frac{dx}{dt}$ are uncorrelated. 
For Gaussian variables, zero correlation is equivalent to independence, and it follows that 
$\langle \frac{dx}{dt}| x \rangle = \langle \frac{dx}{dt} \rangle = 0,$ which parallels 
eq.\ (\ref{eq:dsdtzero}), as derived in \cite{Pan18}.

Similarly, the independence between $x$ and $\frac{dx}{dt}$ gives $\left<\left(\frac{dx}{dt}\right)^2|x \right> = \left<\left(\frac{dx}{dt}\right)^2\right> = \sigma_{dx/dt}^2$. We can thus write, 
\begin{equation}
\frac {\left<\left(\frac{dx}{dt}\right)^2 |x \right>}{ \sigma_{dx/dt}^2} =1.
\label{eq:OUdsdtsq}
\end{equation} 
As noted above, both $\sigma_{dx/dt}^2$ and $ {\left<\left(\frac{dx}{dt}\right)^2 |x \right>}$ are infinite for the Langevin model. On the other hand, for the version with time-correlated noise, $\alpha \Delta t \ll 1,$ we expect that ${\left<\left(\frac{dx}{dt}\right)^2 |x \right>} = \sigma_{dx/dt}^2 = \frac {\beta^2}{\Delta t}$. 

Finally, we calculate $\left<\left(\frac{d^2x}{dt^2}\right)|x \right>$. 
Since $x$ and 
$\frac{d^2x}{dt^2}$ are Gaussian, one may write $\frac{d^2x}{dt^2} = \frac {\langle x \frac{d^2x}{dt^2}\rangle}{\sigma_x^2} x  + \zeta$ where $\zeta$ is independent of $x$. Therefore, we find that $\left<\left(\frac{d^2x}{dt^2}\right)|x \right> = 
 \frac x {\sigma_x^2} {\left<x\frac{d^2x}{dt^2} \right>} = 
- \frac x {\sigma_x^2} {\left<\left(\frac{dx}{dt}\right)^2 \right>}  $,
where the last equality uses the fact that $\left<x \frac{d^2x}{dt^2} \right> = - \left<\left(\frac{dx}{dt}\right)^2 \right>$ in a steady state. In a more convenient form, this gives 
\begin{equation}
\frac { \left<\left(\frac{d^2x}{dt^2}\right)|x \right>}{\left<\left(\frac{dx}{dt}\right)^2 \right>}= -\frac {x}{\sigma_x^2}.
\label{eq:OUdsdt2}
\end{equation}
For the  time-correlated noise model, if $\alpha \Delta t \ll 1$, then $\left<\left(\frac{d^2x}{dt^2}\right)|x \right> = -x \, \sigma^2_{dx/dt}/\sigma_x^2 =
- x \, 2 \alpha / \Delta t$, 
Together eqs.\ (\ref{eq:OUdsdtsq}) and (\ref{eq:OUdsdt2}) provide us with simple expressions to compare against the conditional averages measured from our simulations.  

\subsection*{Acknowledgements}

We would like to acknowledge Blakesley Burkhart, Corentin Cadiou, Javier Castro, Vadim Semenov, and Romain Teyssier for helpful discussions that greatly improved this manuscript. We would also like to thank Edith Falgarone, Pierre Lesaffre, and the organizers of MIST2023 for a stimulating conference that helped us to refine many of the results presented here.  
\subsection*{Funding}
ES and EBII were supported by NASA grant 80NSSC22K1265 and the work includes simulations run on NASA High-End Computing resources. L.P. acknowledges financial support from NSFC under grant No.\ 11973098 and No.\ 12373072. M.B. acknowledges support from the Deutsche Forschungsgemeinschaft under Germany's Excellence Strategy - EXC 2121 "Quantum Universe" - 390833306 and from the BMBF ErUM-Pro grant 05A2023.
\subsection*{Data and materials availability}  All data needed to evaluate the conclusions in the paper are present in the paper and/or the Supplementary Materials.  Additional reduced data sets are available at https://doi.org/10.5061/dryad.zgmsbccn0, and full simulation outputs are available by request from the authors.}

\bibliographystyle{Science}
\bibliography{Turb3.bib}
\vspace{0.2in}


\pagebreak

\renewcommand{\thesection}{S\arabic{section}}
\renewcommand{\thefigure}{S\arabic{figure}}
\renewcommand{\theequation}{S\arabic{equation}}
\renewcommand{\thepage}{S\arabic{page}}
\renewcommand{\thetable}{S\arabic{table}}
\setcounter{section}{0}
\setcounter{figure}{0}
\setcounter{equation}{0}
\setcounter{table}{0}
\setcounter{page}{1}




\section{Supplementary Text}

 \subsection{Accuracy of Conditional Averages}
 \label{sec:accuracyconditional}

\begin{figure}[t]
\begin{center}
\includegraphics[width=0.55\textwidth]{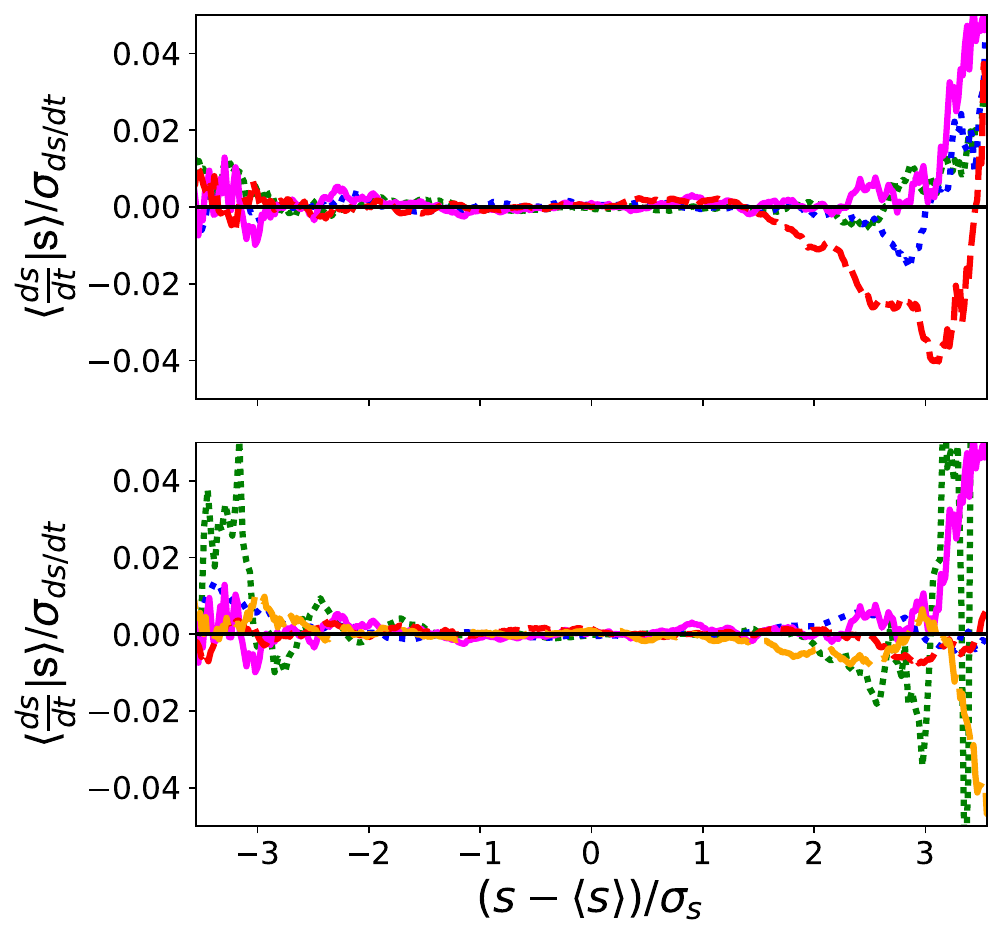}
\end{center}
\vspace{-0.2in}
\caption{{\bf Ratio of the average value of $\frac{ds}{dt}$ as function of $s$ normalized by $\sigma_{{ds}/{dt}}.$ }
The upper panel gives results from the $M_{\rm rms} = 3.2$ simulations with different viscosities and the lower panel gives results from $\nu_0 = 3\times10^{-4} L_{\rm box} c_s$ simulations with different Mach numbers. To aid in comparing the various cases, we shift and rescale the $x$-axis from each run by the mean value and rms of $s.$  Thus for all runs, the curves show the behavior within the range of $\left<s \right> \pm 3.5 \sigma_s.$ The line styles and colors are as in the previous figures.}
\label{fig:Dsdt_norm}
\end{figure}

 \begin{figure}[t]
\begin{center}
\includegraphics[width=0.6\textwidth]{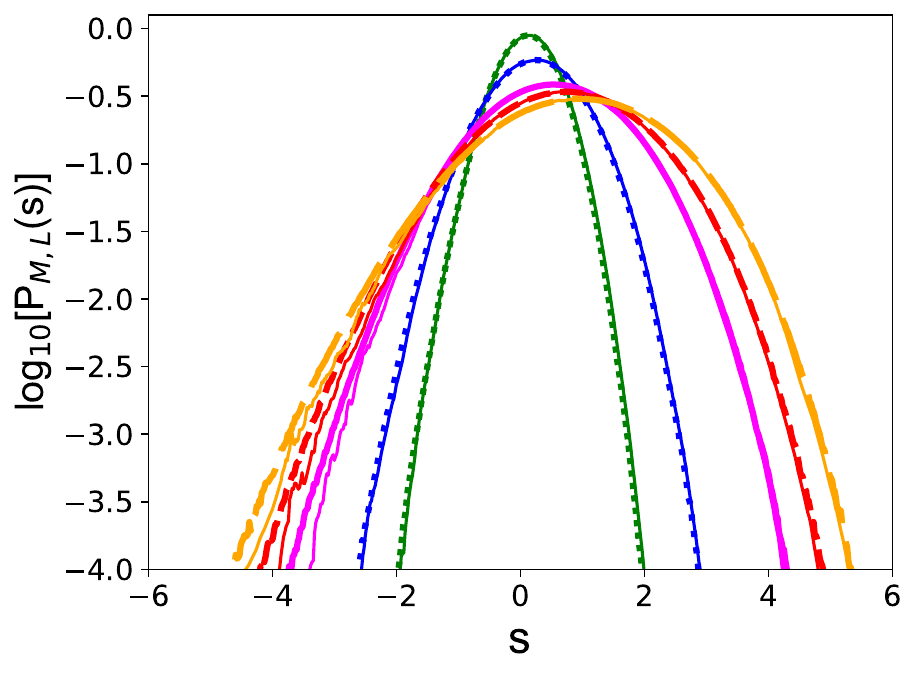}
\vspace{-0.2in}
\end{center}
\caption{{\bf Comparison of the mass weighted probability density function $P_{M,L}(s)$ in the $\nu_0= 3\times10^{-4} L_{\rm box} c_s$ simulations with predictions from the conditional averages. } The colored lines show $P_{M,L}(s)$ for 
M1.3$\nu_0$3 (green dense-dotted), M2$\nu_0$3 (blue dotted), M3$\nu_0$3 (magenta solid), M4$\nu_0$3 (red short-dashed), and M6$\nu_0$3 (orange long-dashed), and the thin solid lines show the prediction given by eq.\ (\ref{eq:psconditional}) based on the ratio of $\left< (\frac{ds}{dt})^2 | s \right>$ and $\left< \frac{d^2s}{dt^2} | s \right>$ as shown in Fig.\ \ref{fig:DsdtsqandR}.  The deviations of the conditional average from the Langevin model at very high densities lead to the nongaussian features of $P_{M,L}(s).$}
\label{fig:PS_cond}.
\end{figure}

 Fig.\ \ref{fig:Dsdt_norm} shows $\left< \frac{ds}{dt} | s \right>$ as computed from our tracer particles, which should be 0 for all $s$ in a steady-state, which serves as a test of the accuracy of our measurements.
These averages were computed by post-processing output files that were written with a cadence of 20 per sound crossing time. In all our simulations the errors are much less than 1\% over the full region from $\left< s \right> -3.5 \sigma_s  \leq s \leq \left< s \right> + 2 \sigma_s$, The errors are also less than 5\% in the $\left< s \right> + 2 \sigma_s \leq s \leq \left< s \right> + 3.5 \sigma_s$ in which there are significantly fewer particles, due in part to the overall skew of $P_{M,L}(s)$ to negative values.  This is in contrast with Eulerian measurements that rely on post-processing, which can give errors of order unity or greater over a similar range \cite{Pan18}. Thus we conclude that our approach provides an accurate measure of the conditional averages of interest.

 \subsection{Confirmation of eq.\ (\ref{eq:psconditional})}
 \label{sec:shape}

 As a confirmation of eq.\ (\ref{eq:psconditional}), in Fig.\ \ref{fig:PS_cond} we replot $P_{M,L}(s),$ comparing the measured values with the predictions from this equation. To minimize the noise in this comparison, we integrate this equation outwards in both directions from the center of the distribution in which the ratio of $\frac{d^2s}{dt^2}$ and $(\frac{ds}{dt})^2$ is best measured.  Here we see that, unlike the Gaussian fits in Fig.\ \ref{fig:PsPdsdt}, this expression provides a good fit to $P_{M,L}(s)$ over the full range of $s$ values. This, in turn, arises from the fact that 
$ \left< \frac{d^2 s}{dt^2} | s \right>/\left< \left(\frac{ds}{dt}\right)^2 | s \right> \approx -({s-\left< s \right>})/{\sigma_s^2}$ as predicted by the Langevin model, with the one key exception this ratio decreases strongly at $s - \left< s \right> \gtrsim 3 \sigma_s,$ indicating a rapid decrease in shock strength at these high densities.

 \subsection{Comparison with Related Literature Measurements}
 \label{sec:postshock}
 
To our knowledge, the Lagrangian PDF of $\frac{ds}{dt}$ and the Eulerian, density-weighted PDF of velocity divergence have not been explored previously.  \cite{Federrath08b} used Lagrangian tracer particles to investigate the mixing process of molecular hydrogen in turbulent clouds. \cite{Price10} and \cite{Konstandin12} compared the PDF of $s$ in supersonic turbulence using both the Eulerian and Lagrangian approaches. \cite{Gotoh93} and \cite{Passot98} examined the PDF of velocity gradient in 1D decaying Burgers turbulence and in highly compressible polytropic turbulence, respectively. \cite{Wang12} studied the volume-weighted PDF of the velocity divergence in a transonic, non-isothermal turbulent flow and found a -5/2 power law distribution for the left wing. \cite{Appel23} studied the 2D volume-weighted PDF of velocity divergence and $s$ in a suite of giant molecular cloud simulations that included gravity, turbulence, stellar feedback, and magnetic fields.  However, due to different simulation settings, one may not obtain a meaningful comparison between these results and the Lagrangian PDF of $\frac{ds}{dt}$  from our simulations. 

It is interesting to contrast our measurements $\xi_s(\tau)$ with the post-shock density profiles measured in \cite{Robertson18} and presented in their Fig.\ 4. In that work, the authors found that the mean density profile behind the shocks was well described as $\rho(x) \propto \exp(-x/h)$ where $x$ is the distance from the shock from and  $h$ the scale height. An exponential atmosphere model is developed to explain the exponential  profile, which also shows that $h$ given by  $\Sigma/\rho_w M_s^2.$, where $\Sigma$ is the swept-up mass  per unit area by the shock that has accumulated in the postshock region, $M_s,$ and $\rho_w$ are the preshock velocity and density respectively. Their finding is a different, but related observation to the one reported here in Fig.\ \ref{fig:s_cor}. 

Our analysis tracks the average behavior of $s=\ln(\rho/\rho_0),$ meaning that an exponential decline in  $\rho/\rho_0$ would appear as a power-law in the variable studied here. Furthermore, our analysis combines all shocks together, without any rescaling as carried out by \cite{Robertson18}. Thus we conclude that our analyses emphasize different features of the turbulent flow, with \cite{Robertson18} describing the physics of the profile behind individual shocks, and our model emphasizing the overall impact of the shock distribution on the evolution of the turbulent medium.

\subsection{Alternative Measures of $\frac{ds}{dt}$}
\label{sec:tracerparticles}
 
FLASH allows for two schemes for interpolating grid quantities onto passive particles, a cloud-in-cell (CIC) mapping that linearly interpolates on a region of one cell size around each particle and a quadratic mapping that calculates a second-order interpolation function using all the adjacent cells. We adopt the CIC mapping throughout this study, but, for comparison, we also included a second set of particles using the quadratic mapping. In this case, for each particle a second-order function of the form $f(x, y, z) = A + B \times (x - x_i) + C \times (x - x_i)^2 + D \times (y - y_j ) + E \times (y - y_j )^2 + F \times (z - z_k) + G \times (z - z_k)^2$ interpolated the grid values onto each particle position $x,y,z$ where the coefficients $A$ through $G$ were fit to the cell containing the particle and those along each face.

\begin{figure}[t]
\begin{center}
\includegraphics[width=1.0\textwidth]{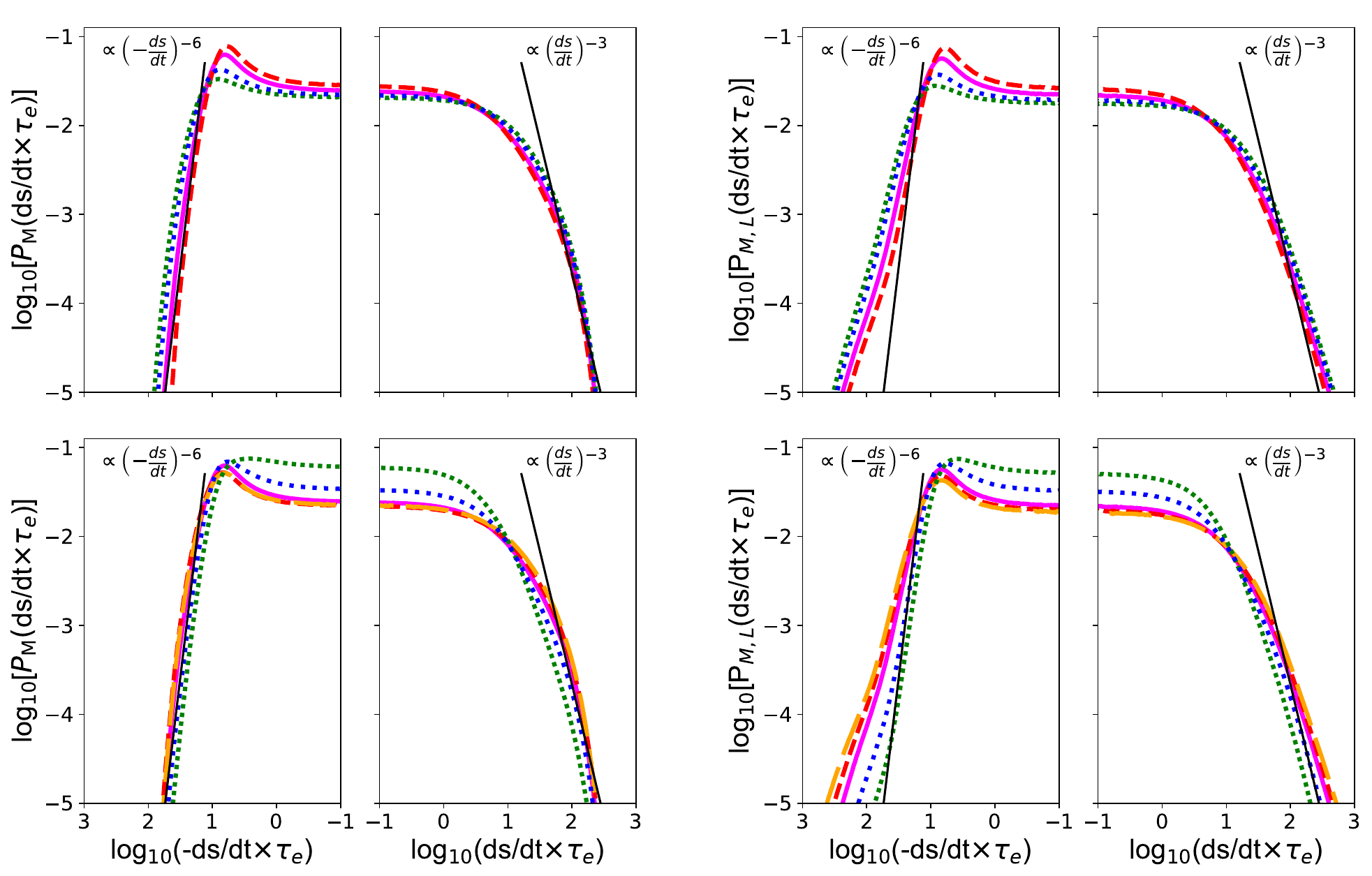}
\vspace{-0.25in}
\end{center}
\caption{{\bf Alternative measures of the probability density function of $\frac{ds}{dt}.$}
 {\em Left:} $P_M(\frac{ds}{dt} \times \tau_e)$ as inferred from the divergence of the velocity on the Eulerian grid. 
 {\em Right:} $P_{M,L}(\frac{ds}{dt} \times \tau_e)$ as inferred from the second-order interpolation scheme. In both rows,
 the upper panel shows the results as a function of viscosity and the lower panel shows the result as a function of Mach number. Colors and line styles match those in previous figures and for ease of comparison the black $(-\frac{ds}{dt})^{-6}$ and $(\frac{ds}{dt})^{-3}$ lines given in this figure are in the same positions as in Fig.\ \ref{fig:PsPdsdt}.}
\label{fig:Pdsdt-rho_VQ}
\end{figure}

In Fig. \ref{fig:Pdsdt-rho_VQ} we show the PDF of $\frac{ds}{dt}$ as computed by two alternate methods to the Lagrangian Cloud-in-cell (CIC) method. In the left panels of Fig.\ \ref{fig:Pdsdt-rho_VQ}, we show the results of inferring the PDF from the Eulerian grid through the use of eq.\ (\ref{eq:dstoV}). Here we see that the overall properties of these measurements are very similar to those inferred from the CIC approach. However, as discussed in \S \ref{sec:pdf} and \ref{sec:conditional}, the Eulerian approach introduces significant errors in $\left<\frac{ds}{dt} \right>$ due to the unresolved $p dV$ work accounted for by the hybrid Riemann solver, an effect that can lead to errors of order unity or greater for $\left<\frac{ds}{dt} | s\right>$ at high $s$ values \cite{Pan18}.

In the right panels of Fig.\ \ref{fig:Pdsdt-rho_VQ}, we show the results of Lagrangian measurements taken from particles in which a quadratic interpolation scheme is used, rather than the CIC approach. As in the CIC case, a two-stage Runge-Kutta scheme is used to integrate forward in time. Unlike the first two approaches, this scheme leads to unphysical tails in the distribution, which we associate with inaccuracies interpolating along the regions immediately in front and behind shocks. For this reason, we avoid applying this approach in deriving our main results.

\subsection{Effect of Numerical Resolution}
\label{sec:resolution}

\begin{figure}[t]
\begin{center}
\includegraphics[width=1.0\textwidth]{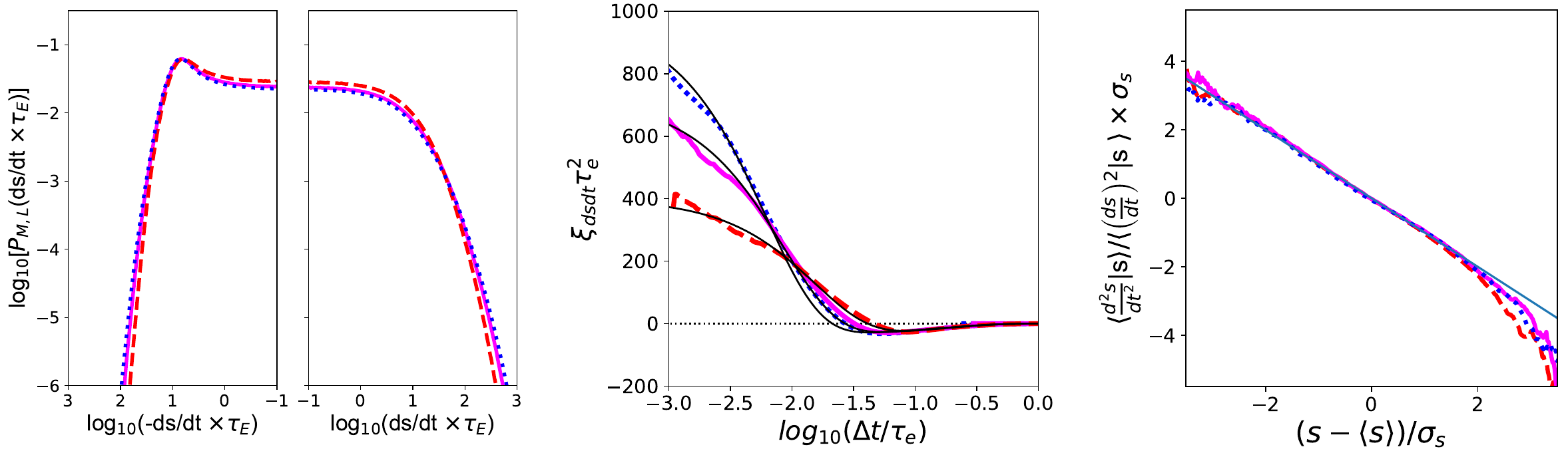}
\vspace{-0.2in}
\end{center}
\caption{{\bf Measurements of the effect of numerical resolution on our simulation results.} Comparison of the results from $M_{\rm rms} = 3.2,$ $\nu_0 = 3.0 \times 10^{-4},$ simulations with $256^3$ cells (red dashed), $512^3$ cells (magenta solid), and $768^3$ cells (blue dotted).  The left panel shows a comparison of $P_{M,L}(ds/dt),$ between the three runs, uncovering a resolution dependence of the tails due to changes in effective viscosity.  The central panel shows the correlation function of $\frac{ds}{dt}$ between the three runs, as compared to analytic models with $\Delta t_{\rm shock}$ decreasing with resolution. Finally, the rightmost panel shows  $\left<\left(\frac{d^2s}{dt^2}\right)|s \right>/\left<\left(\frac{ds}{dt}\right)^2|s \right>$ across these runs.  This shows that the linear behavior of this ratio as a function of $s$ and the downturn at dense regions are largely independent of resolution.}
\label{fig:converge}.
\end{figure}

To test the effects of numerical resolution on our results,  we have carried out two additional simulations, with one with $256^3$ cells and $128^8$ particles, and a second with $768^3$ cells and $192^3$ particles.  In both cases, we adopted the same driving as our fiducial M3$\nu_0$3 and an explicit viscosity of  $\nu_0 = 0.0003 L_{\rm box} c_s.$ The results of these simulations are shown in Fig.\ \ref{fig:converge}, and their properties are summarized in Table \ref{tab:converge}. Computing the effective viscosity of these runs as in \S \ref{sec:effectivenu} shows that $\nu$ is dominated by numerical effects in the $256^3$ case, while in $768^3$ case the explicit viscosity provides the largest contribution to $\nu$.

\begin{table*}[t]
 \centering
 \resizebox{1.0\textwidth}{!}{%
\begin{tabular}{|l|c|c|c|c|c|c|c|c|c|c|}
\hline
\, \, Name  & $\nu$  & \, Re \,  &$ \left< s \right>$ & $\sigma_s^2$ & $\mu_s$ & $\sigma^2_{ds/dt}  \tau_e^2$ & $\sigma_{{ds}/{dt}^+}^2$  &  $\alpha \tau_e$ & $\beta^2 \tau_e$ & $\Delta t_{\rm shock}/\tau_e$ \\
\, \,     & $(L_{\rm box} c_s)$ & &  & & & & $/\sigma_{{ds}/{dt}}^{2}$  & & & \\
\hline
M3$\nu_0$3\_256      & $10 \times 10^{-4}$   &1,600 & 0.54  & 1.05 & -0.092 & 400 & 0.84 & 6.0 & 13.8 & 0.0316   \\
M3$\nu_0$3              & $6.1 \times 10^{-4}$   & 2,600  & 0.52  & 1.03 & -0.063 & 720 & 0.88 & 6.0 & 13.1 & 0.0173   \\
M3$\nu_0$3\_718    & $5.0 \times 10^{-4}$   & 3,200  & 0.52  & 1.02 & -0.108 & 980&  0.90 & 6.0 & 12.7 & 0.0125  \\
\hline 
\end{tabular}}
\vspace{0.1in}
\caption{Properties of our convergence test simulations. For all runs the rms and average mass-weighted Mach number,  eddy turnover time, and explicit viscosity are $M_{\rm rms}= 3.2$, $M_{\rm ave} = 2.8$, $\tau_e = 0.16 \tau_{\rm sc},$ and $\nu_0 = 3.0 \times 10^{-4} L_{\rm box} c_,$ respectively.  Columns show the run name, the average time step, the effective viscosity, the effective Reynolds number, the mean, variance, and skewness of $P_{M,L}(s)$, the variance of $P_{M,L}(\frac{ds}{dt})$ and the fraction of the variance due to compressions,  and the fit parameters of our random walk models with time-correlated noise. }
\label{tab:converge}
\end{table*}%

In the left panel of Fig.\ \ref{fig:converge}, we see that at $256^3$ resolution, the tails of the PDF shrink slightly, due to the large effective viscosity.  Similarly, with $768^3$ cells, the tails of the PDF move outwards slightly, as shocks and expansions are slightly better resolved. As discussed in \S \ref{sec:tc}, this suggests that the thickness of the shocks is not fully resolved in the fiducial simulations.  

We quantify this further in the central panel of Fig.\ \ref{fig:converge}, which shows $\xi_{ds/dt}(\tau)$ for each of these runs.  Like the $512^3$ runs with different explicit viscosities discussed above, changing resolution has a significant impact on the short-time structure of the correlation function, while having a minimal impact on the long-time behavior and the overall change in $s$ across a typical shock.  Thus, as shown in Table \ref{tab:converge},  while  $\sigma^2_{ds/dt}  \tau_e^2$ systematically increases with resolution, $\sigma_s^2,$ is largely unchanged across these run.  The overall fraction of the variance of $\frac{ds}{dt}$ contributed by compressions,  $\sigma^2_{ds/dt+} /  \sigma^2_{ds/dt},$ is also similar across the runs.

In the central panel of \ref{fig:converge}, we also show the results of correlated-noise models fits to each of these runs, with parameters given in Table \ref{tab:converge}. This shows that the long-time behavior, as quantified by $\alpha,$ and the degree of stochastic driving, as quantified by $\beta$ are largely resolution independent, while $\Delta t_{\rm shock}$ decreases significantly with better resolution.  

Finally, the right panel of Fig.\ \ref{fig:converge}, shows the ratio of $\left<\left(\frac{d^2s}{dt^2}\right)|s \right>$ and $\left<\left(\frac{ds}{dt}\right)^2|s \right>,$ which determines the functional form of the $P_M(s)$ as given by eq.\  (\ref{eq:shape}). Here we see that the linear behavior of this ratio as a function of $s,$ which quantifies the acceleration/deceleration of shocks due to density gradients is independent of resolution.  Similarly, the downturn at high $s$ values, which is likely to due thermal-pressure exceeding ram pressure, is largely the same across the runs.  These features, in turn, lead to  $P_M(s)$ that is nearly Gaussian, with an overall variance and significant negative skewness that are largely independent of resolution, as quantified in Table \ref{tab:converge}.

\end{document}